\documentclass[12pt]{article}
\usepackage{authblk}
\usepackage[margin=1.2in]{geometry}
\usepackage{graphicx}
\usepackage{amssymb}
\usepackage{xfrac}
\usepackage{cite}
\begin{document}

\newif\ifarxiv
\arxivtrue

\newcommand{\bra}[1]{\langle #1|}
\newcommand{\ket}[1]{|#1\rangle}
\newcommand{\braket}[2]{\langle #1|#2\rangle}
\newcommand{\brag}{\langle\langle}
\newcommand{\ketg}{\rangle\rangle}
\newcommand{\s}{\sigma}
\newcommand{\clebsch}[3]{\langle #1;#2 \vert #3\rangle}
\newcommand{\expect}[1]{\langle #1 \rangle}
\newcommand{\e}{\varepsilon}
\newcommand{\w}{\omega}
\newcommand{\G}{\Gamma}
\newcommand{\up}{\uparrow}
\newcommand{\down}{\downarrow}

\newcommand{\beq}{\begin{eqnarray}}
\newcommand{\eeq}{\end{eqnarray}}
\newcommand{\be}{\begin{equation}}
\newcommand{\ee}{\end{equation}}
\newcommand{\de}{{\rm d }}
\newcommand{\dd}{\partial}
\newcommand{\h}{\hslash}

\title{\Large \bf Observation of the SU(4) Kondo state in a\linebreak double quantum dot}
\author[1]{\normalsize A.~J.~Keller}
\author[1,$\dag$]{S.~Amasha}
\author[2]{I.~Weymann}
\author[3,4]{C.~P.~Moca}
\author[1,$\ddag$]{I.~G.~Rau}
\author[5]{J.~A.~Katine}
\author[6]{Hadas~Shtrikman}
\author[3]{G.~Zar\'{a}nd}
\author[1,*]{D.~Goldhaber-Gordon}
\affil[1]{\footnotesize Geballe Laboratory for Advanced Materials, Stanford University, Stanford, CA 94305, USA}
\affil[2]{Faculty of Physics, Adam Mickiewicz University, Pozna\'n, Poland}
\affil[3]{BME-MTA Exotic Quantum Phases ``Lend\"{u}let'' Group, Institute of Physics, Budapest University of Technology and Economics, H-1521 Budapest, Hungary}
\affil[4]{Department of Physics, University of Oradea, 410087, Romania}
\affil[5]{HGST, San Jose, CA 95135, USA}
\affil[6]{Department of Condensed Matter Physics, Weizmann Institute of Science, Rehovot 96100, Israel}
\affil[$\dag$]{Present address: MIT Lincoln Laboratory, Lexington, MA 02420, USA}
\affil[$\ddag$]{Present address: IBM Research -- Almaden, San Jose, CA 95120, USA}
\affil[*]{Corresponding author; goldhaber-gordon@stanford.edu}
\date{}
\maketitle

{\bf Central to condensed matter physics are quantum impurity models, which describe how a local degree of freedom interacts with a continuum. Surprisingly, these models are often universal in that they can quantitatively describe many outwardly unrelated physical systems. Here we develop a double quantum dot-based experimental realization of the SU(4) Kondo model, which describes the maximally symmetric screening of a local four-fold degeneracy. As demonstrated through transport measurements and detailed numerical renormalization group calculations, our device affords exquisite control over orbital and spin physics. Because the two quantum dots are coupled only capacitively, we can achieve orbital state- or ``pseudospin''-resolved bias spectroscopy, providing intimate access to the interplay of spin and orbital Kondo effects. This cannot be achieved in the few other systems realizing the SU(4) Kondo state.}

\newpage

Kondo physics is at the heart of heavy fermion materials and heavy fermion superconductivity \cite{Hewson2003,Coleman2007},
underpins Kondo insulators like samarium hexaboride \cite{Zhang2013,Cooley1995,Menth1969,Dzero2010,Botimer2012,Wolgast2012}, and provides a path toward realizing non-Fermi liquids \cite{Potok2007,CoxZawadowski}. In the simplest version of the Kondo effect, itinerant electrons screen a local spin-1/2 moment through virtual spin-flip processes, yielding a many-body spin singlet state. The observation of this behavior in semiconductor quantum dots and subsequent confirmation of universal scaling \cite{Goldhaber-GordonNat1998,Goldhaber-GordonPRL1998,Cronenwett1998} ignited a surge of interest in studying Kondo physics using mesoscopic or nanoscale systems, where key parameters may be tuned in situ.

Many insights have been gained by studying the Kondo effect in systems as diverse as carbon nanotubes \cite{Nygard2000}, complex oxide surfaces \cite{Lee2011}, nanowires \cite{Kretinin2011,Jespersen2006}, magnetic adatoms on metallic surfaces \cite{Madhavan1998,Li1998,Otte2008}, vertical quantum dots \cite{Sasaki2004}, and break junctions \cite{Parks2007}. Lithographically-defined quantum dots in GaAs/AlGaAs heterostructures complement these studies by providing a platform for designing quantum impurity systems with particular degeneracies and interactions \cite{Potok2007}, rather than relying on those intrinsic to a particular material. Specifically, quantum dots should enable studies of SU(4)-symmetric Kondo effect \cite{Zarand2003, Borda2003, LeHur2007,Lopez2005,Sato2005,Eto2005}, which are relevant to systems that possess not only spin but also valley or orbital degrees of freedom, like carbon nanotubes \cite{Choi2005,Anders2008,Jarillo-Herrero2005,MakarovskiPRB2007,MakarovskiPRL2007,Delattre2009} or silicon field effect transistors \cite{Tettamanzi2012, Lansbergen2010}.

In this article, we study experimentally and theoretically the transport properties of a double quantum dot (DQD) with strong interdot capacitive coupling, with each dot tunnel-coupled to its own pair of leads. This device geometry allows for unprecedented control of the orbital degrees of freedom. We find excellent agreement between our NRG calculations and the experimental data over a wide range of gate voltages and temperatures, enabling us to demonstrate the SU(4) Kondo state for the first time in a double quantum dot, and to identify universal SU(4) Kondo scaling. Furthermore, the unique orbital state- or ``pseudospin''-resolution of our device \cite{Amasha2013} allows us to explore the orbital structure of the SU(4) Kondo state and study how simultaneous Zeeman and pseudo-Zeeman fields manifest differently in each pseudospin channel.

\ifarxiv
\begin{figure}
\ifarxiv
\includegraphics[width=6in]{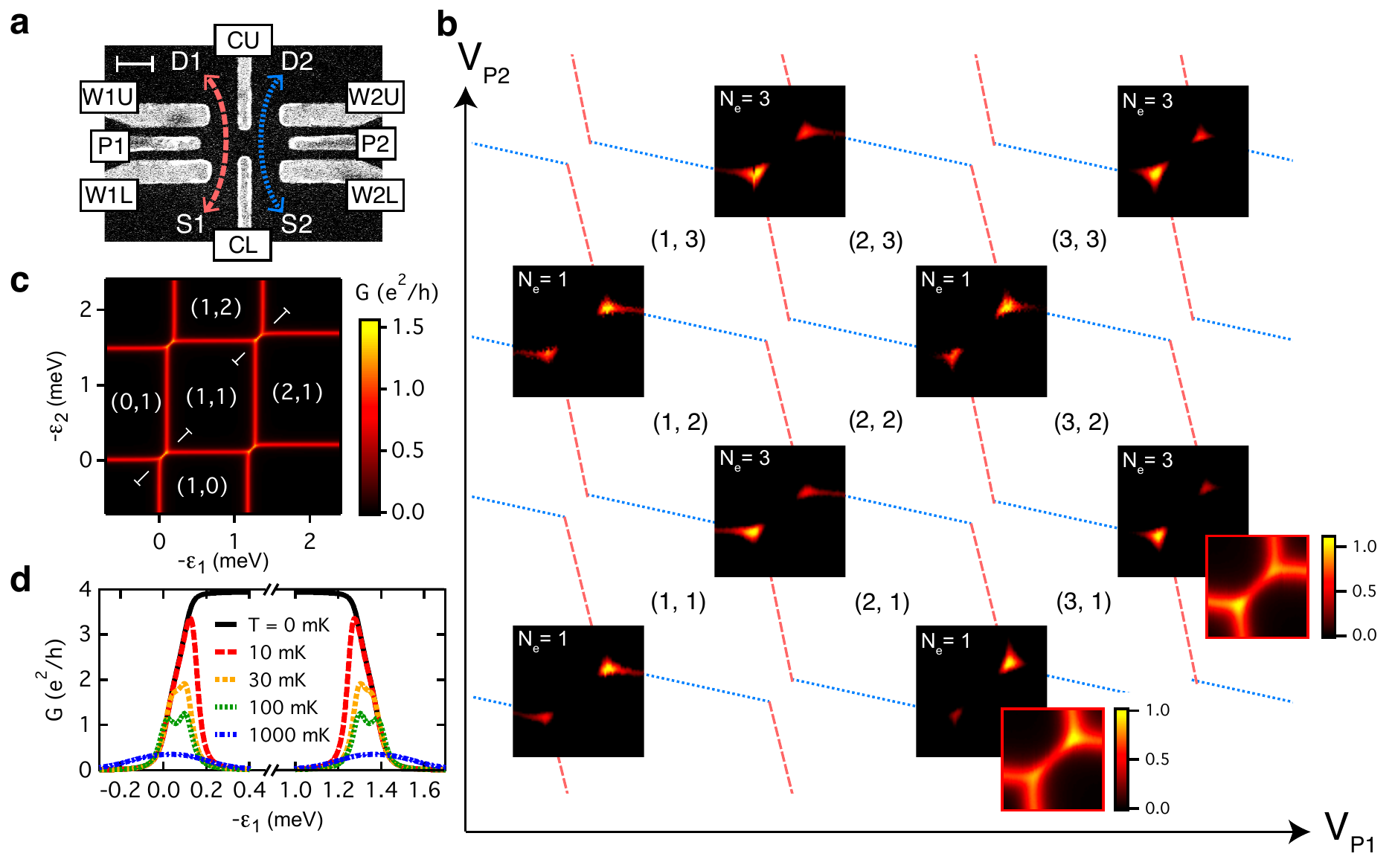}
\fi
\caption{\small {\bf Survey of conductance at orbital degeneracies.}
{\bf (a)} SEM micrograph of a double quantum dot like the one measured.  The scale bar (upper left) is 200nm. Currents are measured from S1 to D1 (red dashed arrows) and from S2 to D2 (blue dotted arrows).
{\bf (b)} Experimental conductance $G$ = $G_1$ + $G_2$ at $N_e = 1$ and $N_e = 3$ LBTPs. Electron occupancies are labeled within each hexagon, relative to a (0,0) hexagon where each dot's occupancy is only known modulo 2. Each square of data corresponds to a region spanning 3 mV in $V_{P1}$ and $V_{P2}$. The color scales are individually set so that only data between 75\%--100\% of the maximum conductance are visible. Two examples of the data with the color scale unsaturated are also shown (bottom-right). The red lines and blue lines behind the data schematically depict the charge stability diagram (not to scale). Red dashed (blue dotted) lines indicate Coulomb blockade peaks for dot 1 (2). 
{\bf (c)} Charge stability diagram from NRG calculations of $G$ = $G_1$ + $G_2$. The parameters used were $T = 30$~mK, $B =$~0, $U_1 = 1.2$~meV, $U_2 = 1.5$~meV, $U^\prime = 0.1$~meV, $\Delta_1 = 0.017$~meV, $\Delta_2 = 0.0148$~meV, $\alpha_1 = \alpha_2 = 1$.  $\epsilon_{1(2)}$ is the energy of the dot 1(2) state above the Fermi energies of the leads. The white text labels the number of electrons on dots 1 and 2. The white brackets mark the location of $N_e = 1$ (lower-left) and $N_e = 3$ (upper-right) LBTPs.
{\bf (d)} Cuts along the $N_e = 1$ and $N_e = 3$ LBTPs of (c) at zero and finite temperatures, parameterized by $-\epsilon_1$. The left part of (d) corresponds to $N_e = 1$ and the right part corresponds to $N_e = 3$. At 30 mK, the conductance is higher at the peaks nearest to the (1,1) hexagon, as in the experimental data of (b).
\label{fig1}
}
\end{figure}

\fi

Our DQD, formed by lithographically-defined gate electrodes (Fig. 1a), resides in the two-dimensional electron gas (2DEG) of a GaAs/AlGaAs heterostructure with mobility $2 \times 10^6$ cm$^2$/Vs and electron density $2 \times 10^{11}$ cm$^{-2}$. The electrochemical potential of each dot is tuned with its respective P gate, and the dot-lead tunnel rates are tuned with the W gates. The interdot tunnel rates are tuned to be negligible using the C gates. Differential conductances $G_1$ and $G_2$ are measured through dots 1 and 2 by applying independent 1 $\mu V$ ac excitations at the source terminals S1 and S2 (f = 85 and 102 Hz) and measuring the resulting currents through drain terminals D1 and D2. 

A DQD may be modeled by charging energies, tunnel couplings, and effective dot levels. The charging energy $U_{1(2)}$ is the interaction energy of two electrons on dot 1 (2). The capacitive coupling between the dots results in an interdot charging energy $U^\prime$, the energy by which states in dot 2 increase if an electron is added to dot 1, or vice versa. Each of the dots is tunnel-coupled to a pair of leads, causing the discrete states of each dot to hybridize with those leads. The energy scale of this hybridization, $\Delta_{1(2)}$ for dot 1 (2), would be the linewidth of a state in the absence of Coulomb interaction. At a given temperature $T$, the experimentally measured linewidth $\Gamma_{1(2)}$ is related to $\Delta_{1(2)}$ by a universal function. The coupling of a state to the source and drain leads may be asymmetric, encapsulated in the asymmetry factor $\alpha_{1} = 4\Delta_{1S} \Delta_{1D}/(\Delta_{1S}+\Delta_{1D})^2$ for dot 1, and likewise for dot 2. Finally, the dot levels $\epsilon_{1,2}$ can be viewed as energy gained by transferring an electron to the empty quantum dots. Most of these model parameters may be explicitly determined or inferred from experimental data with the aid of theory (supplemental info).

The SU(4) Kondo effect is only expected for particular parameter regimes. As a function of $\epsilon_1$ and $\epsilon_2$, the summed experimental conductance $G \equiv G_1 + G_2$ through the DQD exhibits a hexagonal ``honeycomb'' charge stability diagram  \cite{vanDerWiel2002}. Within each hexagon, the electron occupancies of dot 1 and 2 are integers $(N_1, N_2)$. Were $U^\prime = 0$, the charge configurations $(N_1, N_2)$, $(N_1+1, N_2)$, $(N_1, N_2+1)$, $(N_1+1, N_2+1)$ could all be degenerate. The interdot capacitance breaks this degeneracy, resulting in a pair of triple points where three of these orbital configurations are degenerate. Along a line between triple points (``LBTP''), two orbital configurations are degenerate, $(N_1+1, N_2)$ and $(N_1, N_2+1)$. This degeneracy constitutes a pseudospin, and along this line the pseudo-Zeeman splitting $E_{PZ}$ equals zero. If the tunnel coupling is tuned to be weak, Kondo-enhanced conductance is seen along the LBTP but not elsewhere in the charging hexagons, underscoring the importance of the pseudospin degeneracy \cite{Amasha2013,Hubel2008}. We will only consider (even,odd)/(odd,even) degeneracies, where SU(4) Kondo is expected \cite{Borda2003}. The (even,even)/(odd,odd) LBTPs may also exhibit related phenomena \cite{Okazaki2011,Busser2012}, but will not be considered further here.

Along the (even,odd)/(odd,even) LBTPs, we observe experimentally a subtle pattern in conductance (Fig. 1b), consistent with particle-hole symmetry for a four-fold degenerate state, where the four-fold degeneracy is established by the spin and pseudospin degrees of freedom. The color scales have been saturated to emphasize that the conductance is always highest near the (odd, odd) hexagon, where quantum fluctuations are stronger due to the high internal spin degeneracy of the excited (odd,odd) state. For these measurements, $\Gamma_1 = \Gamma_{1S} + \Gamma_{1D}$ and $\Gamma_2 = \Gamma_{2S} + \Gamma_{2D}$ have been tuned to $\sim 0.03$ meV, with nearly symmetric source-drain coupling. The pattern is robust against small source-drain biases (supplemental info).

To understand this pattern, consider the (0,1)/(1,0) LBTP (bottom-left of Fig. 1b). The unpaired electron tunneling out of the double dot followed by an electron tunneling back in from the leads can flip the spin, the pseudospin, or both simultaneously. In this sense, all four degenerate states are equivalent, and the Kondo screening of the combined pseudospin and spin degeneracy is described by the SU(4) symmetry \cite{Note}. The conductance enhancement along the (0,1)/(1,0) LBTP may be termed ``\sfrac{1}{4}-filling'' SU(4) Kondo effect in that a four-fold degeneracy of the double dot is filled by only one electron. Because of the particle-hole symmetry of the four-fold degenerate state, the (2,1)/(1,2) LBTPs also exhibit equivalent \sfrac{1}{4}-filling SU(4) Kondo, but the impurity is hole-like. We label the LBTPs $N_e = 1$ for an electron-like impurity or $N_e = 3$ for a hole-like impurity. This number does not denote the total electron occupation (modulo 4), as adding two electrons to either dot results in the same type of LBTP. This contrasts with carbon nanotubes, which exhibit more conventional four-electron shell filling. In a DQD, because the four-fold degeneracy is strongly broken away from an LBTP, lower energy electrons pair off into two-electron singlet states on each dot and may be largely ignored. 

NRG calculations (details in supplemental info) of the summed conductance $G = G_1 + G_2$ through the DQD (Fig. 1c), computed for realistic device parameters, support this interpretation. The sign of the axes corresponds directly with the experimental gate voltages. Because Fig. 1c is calculated at a finite temperature of 30~mK, the conductance in the (1,1) valley between the $N_e = 1$ (bottom-left) and $N_e = 3$ (upper-right) LBTPs is small, as is also evident from calculations of cuts along the LBTPs (Fig. 1d). Note that the calculated $G \approx 4e^2/h$ conductance at $T = 0$ (Fig. 1d) is from SU(2) Kondo rather than SU(4) Kondo; since $U^\prime \ll U_1,U_2$, the ``\sfrac{1}{2}-filling'' SU(4) Kondo effect in the (1,1) valley is not realized \cite{Galpin2006, Galpin2005}. Instead, an ordinary SU(2) spin Kondo effect occurs in each dot. This spin Kondo effect is characterized by a very small Kondo temperature $T_K$ such that even a small temperature $T$ (e.g. 10~mK in Fig. 1d) is enough to completely suppress the conductance. In contrast, the higher degeneracy in the SU(4) Kondo effect leads to higher Kondo temperatures along the $N_e = 1$ and $N_e = 3$ LBTPs. This survey of LBTPs alone does not confirm SU(4) Kondo, but the elegant four-fold pattern is suggestive and motivates further investigations.

The experimental search for SU(4) conductance scaling is not straightforward. In a typical quantum dot Kondo system, by going to a temperature $T\gg T_K$, one can determine $\epsilon$ from the position of the bare Coulomb blockade resonances \cite{Goldhaber-GordonPRL1998}. In this system, $T$ cannot be increased much beyond $T_K$ before $k_B T$ becomes comparable to $U^\prime = 0.1$~meV, hindering experimental identification of the bare resonances. To determine $\epsilon_1$ and $\epsilon_2$ experimentally requires first establishing agreement with NRG calculations, where $\epsilon_1$ and $\epsilon_2$ are given. We proceed to examine the temperature-dependent conductance scaling near the point of maximum symmetry, $\epsilon_1 = \epsilon_2 = -U^\prime/2 = -0.05$~meV, where for symmetrical dots NRG predicts the summed conductance to approach $2e^2/h$ as $T\rightarrow 0$. Because the entire LBTP is expected to exhibit SU(4) scaling, for our comparison between theory and experiment we will work slightly away from the point of maximum symmetry to attain higher $T_K$.

\ifarxiv
\begin{figure}
\ifarxiv
\includegraphics[width=6in]{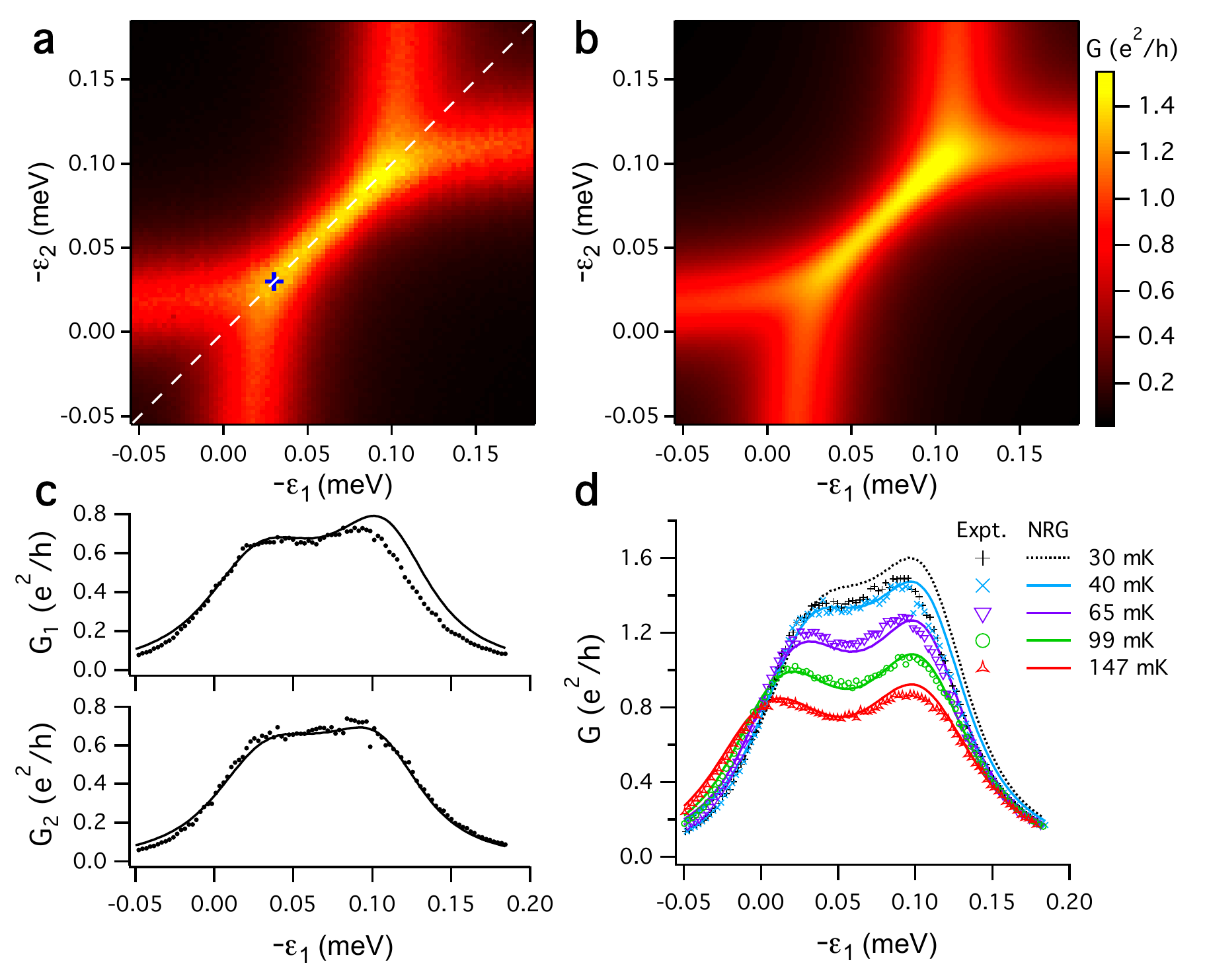}
\fi
\caption{\small {\bf Agreement between experimental data and NRG calculations.}
Experimentally measured {\bf (a)} and theoretically computed {\bf (b)} total conductance at an $N_e=1$ LBTP in units of $e^2/h$ at $T = 40$~mK. The white dashed line corresponds to zero detuning. The blue crosses mark the point $\epsilon_1 = \epsilon_2 = -0.03$~meV, where the SU(4) scaling will be demonstrated in Fig. 3. The parameters used in the NRG calculations are the same as in Fig. 1c and 1d, except with $\alpha_1 = \alpha_2 = 0.875$. 
{\bf (c)} Conductances $G_1$ (top) and $G_2$ (bottom) along the LBTP, indicated by the dashed lines in panels (a) and (b). Dots denote the experimental data, solid lines are the NRG results.  
{\bf (d)} Total conductance along the LBTP for a subset of measured temperatures. Symbols denote the experimental data, solid lines are the NRG results. The agreement between the experiment and theory over the range of energies shown is typical for $T \ge 40$~mK, but the experimental data saturate below 40~mK.
\label{fig2}
}
\end{figure}
\fi

At 40~mK, summed zero-bias conductance from experiment (Fig. 2a) and NRG calculations (Fig. 2b) agree excellently over a range of $\epsilon_1$ and $\epsilon_2$ encompassing the entire LBTP. To measure experimentally along the $-\epsilon_1$ and $-\epsilon_2$ axes, voltages $V_{P1}$ and $V_{P2}$ were swept simultaneously. This was necessary to compensate for the finite cross-capacitance between gate P1 and dot 2, and vice versa. The experimental axes were scaled into units of energy by using bias spectroscopy to determine the couplings of P1 and P2 to the energy levels of the two dots.  For the calculation of Fig. 2b, only the parameters $\alpha_1$ and $\alpha_2$ should be considered free parameters; here $\alpha_1=\alpha_2=0.875$. 

The white dashed line in Fig. 2a corresponds to keeping the pseudo-Zeeman splitting $E_{PZ}$ equal to zero. Fig. 2c shows pseudospin-resolved conductances $G_1$ and $G_2$ along the $E_{PZ} = 0$ line, parameterized by $\epsilon_1$. Reasonable agreement between theory (solid lines) and experiment (points) is attained in both channels, especially in view of the extreme sensitivity of $G_1$ and $G_2$ to the precise cut direction \cite{footnote1}. Using the same NRG parameters, the summed conductance from theory and experiment agree over a range of temperatures at least up to 150~mK (Fig. 2d), although the 22~mK (omitted for clarity) and 30~mK experimental data saturate without reaching their expected low-temperature limits. We defer discussion of the saturation to later in the paper. Nonetheless, the excellent agreement over a range of temperatures and energies, even in the pseudospin-resolved conductances, allows us to use theory to identify the maximum symmetry point in the experimental data.

In Fig. 3, the summed experimental conductance $G$ at $\epsilon_1 = \epsilon_2 = -0.03$ meV is compared with NRG calculations, as well as with the universal SU(4) and SU(2) scaling functions. The Kondo temperatures $T_{K SU(2)}$ and $T_{K SU(4)}$ for the scaling functions have been chosen to provide best fits to the experimental data. The experimental data (open circles) are described well by either the SU(4) scaling function (blue dash-dotted line) or the NRG calculations (black solid line), whereas the SU(2) scaling function (red dashed line) does not provide a good description. The point $\epsilon_1 = \epsilon_2 = -0.03$ meV was chosen instead of the maximum symmetry point $-U^\prime/2$ ($-0.05$~meV) because $T_K$ is larger at $-0.03$~meV, allowing us to experimentally probe conductance closer to the low-temperature limit. This is an important consideration given that the experimental conductance empirically saturates at $T = 40$~mK throughout the LBTP (Fig. 2d). The point $\epsilon_1 = \epsilon_2 = -0.03$~meV with elevated $T_K$ is nearer to the (0,0) end of the LBTP than to the (1,1) end where pure spin fluctuations should be suppressed. However, even at $\epsilon_1 = \epsilon_2 = -0.04$ or $-0.05$~meV, the data are consistent with the NRG calculations and SU(4) scaling except at or below 40--45 mK. This is in keeping with the expectation that SU(4) scaling should hold along the entire LBTP. Note that simply fitting to the popular empirical Kondo form \cite{Goldhaber-GordonPRL1998} would have missed this saturation (supplemental info). We conclude that strong conductance enhancements along the $N_e = 1$ LBTP are due to the SU(4) Kondo effect, and the SU(4) Kondo state must also appear at the $N_e = 3$ LBTP due to particle-hole symmetry, as expected unambiguously in theory and suggested by the data of Fig. 1.

\ifarxiv
\begin{figure}
\ifarxiv
\includegraphics[width=6in]{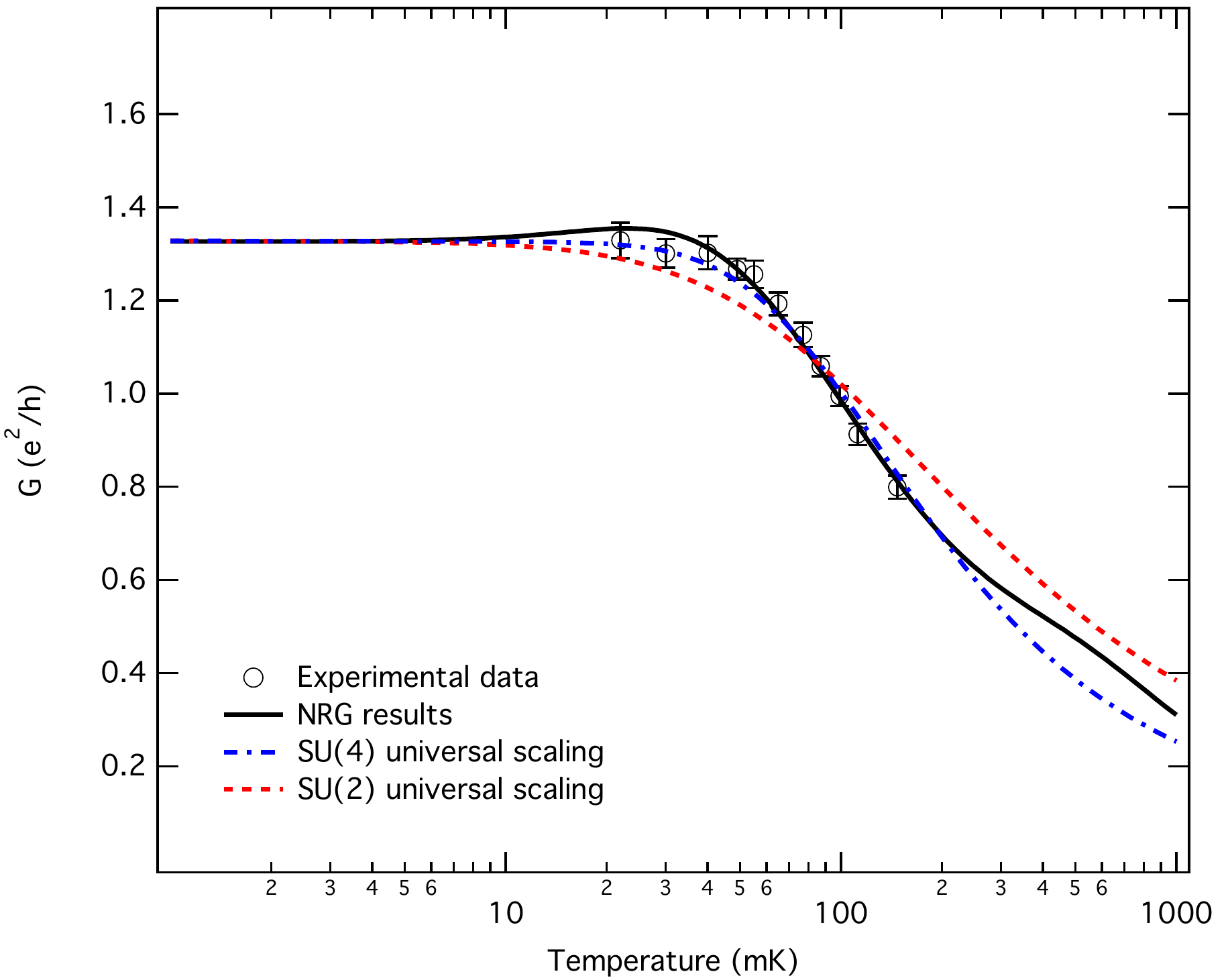}
\fi
\caption{\small {\bf Universal scaling of the conductance.}
Experimental data for the temperature dependence of the conductance (symbols) 
at $\epsilon_1 = \epsilon_2 = -0.03$~meV in Fig. 2d. Experimental data are compared with NRG results as well as with the universal SU(4) and SU(2) curves using best-fit Kondo temperatures $T_{K SU(2)} = 320$~mK and $T_{K SU(4)} = 220$~mK.
\label{fig3}
}
\end{figure}
\fi

We now consider using the DQD to perform pseudospin-resolved bias spectroscopy of the SU(4) Kondo effect. Experimentally, we can perturb the Kondo ground state by breaking either the pseudospin or the spin degeneracy. A pseudo-Zeeman splitting $E_{PZ}$ is achieved with gate voltage-controlled detuning of the orbital states, whereas a Zeeman spin splitting $E_Z$ is achieved by applying a magnetic field precisely in the plane of the heterostructure using a two-axis vector magnet. For simplicity we fix $E_Z$ and smoothly vary $E_{PZ}$.

Fig. 4 shows experimentally-measured conductances $G_1$ (Fig. 4b) and $G_2$ (Fig. 4d) as a function of source-drain biases $V_{SD1}$ = $V_{SD2}$, alongside pseudospin-resolved spectral functions $A_1$ (Fig. 4c) and $A_2$ (Fig. 4e) calculated via NRG. The data were taken in a 1.0 T Zeeman field at an $N_e = 3$ LBTP. We determine $E_Z \equiv |g|\mu_{B}B =$ 0.025 meV for $B = 1.0$ T using a measured g-factor $|g| = 0.44$, in agreement with $g=-0.44$ for GaAs (supplemental info).  The parameters used for the NRG calculation are essentially the same as in previous figures, but describe the $N_e = 3$ LBTP by employing a particle-hole transformation. A peak in either spectral function corresponds to the opening of an inelastic channel. Peaks at positive (negative) energy mark when the energy of incoming electrons (holes) matches a state reachable through an exchange process. The spectral functions should describe the bias spectroscopy up to constants of proportionality, neglecting decoherence, finite level spacing, and nonequilibrium effects (minimized by using asymmetric coupling of source and drain). For simplicity, we will refer to the horizontal axes $\omega$ and $-eV_{SD}$ interchangeably.

\ifarxiv
\begin{figure}
\ifarxiv
\includegraphics[width=6in]{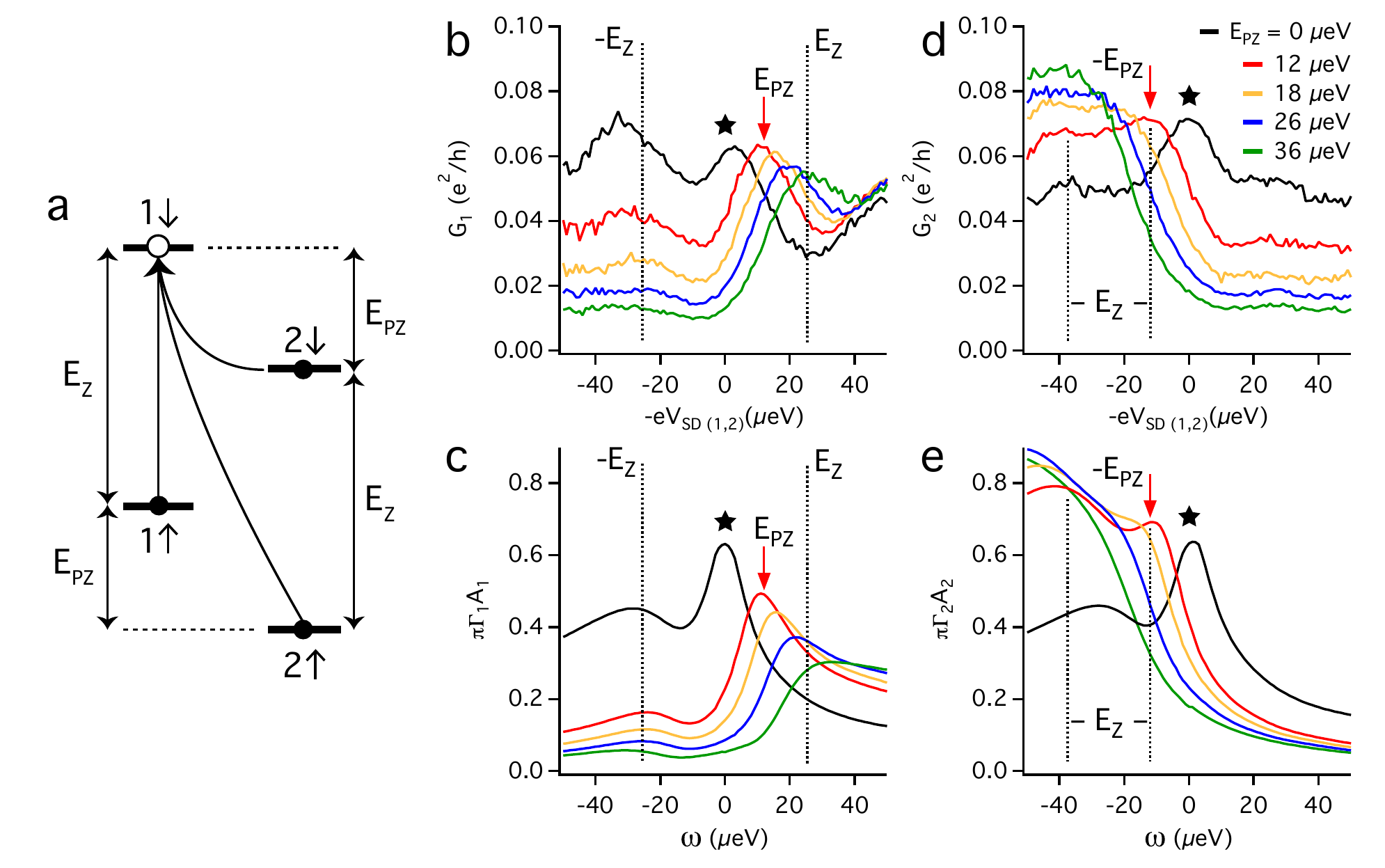}
\fi
\caption{\small {\bf Orbital state-resolved bias spectroscopy of the SU(4) Kondo resonance.}
{\bf (a)} Inelastic transitions between Zeeman-split states of dot 1 and dot 2 at an $N_e = 3$ LBTP.
{\bf (b)} Experimental conductance $G_1$ for dot 1 in a 1.0 T Zeeman field.  The five traces correspond to different values of $E_{PZ}$, with $E_{PZ}>0$ meaning dot 1 is favored to hold the unpaired electron. A predicted ``spin-Kondo like'' feature at $-eV_{SD} \approx -E_Z$ is observed. The star marks a purely orbital Kondo state for $E_{PZ} = 0$ which appears in both dots.
{\bf (c)} Calculated spectral function $A_1$ for dot 1 in a 1.0 T Zeeman field. 
{\bf (d)} Experimental conductance $G_2$ for dot 2 in a 1.0 T Zeeman field. For the same increments in $E_{PZ}$, no peaks are pinned to a particular energy, and instead they move with the pseudo-Zeeman splitting.
{\bf (e)} Calculated spectral function $A_2$ for dot 2 in a 1.0 T Zeeman field. 
For all panels, $\Gamma_1, \Gamma_2 \approx 0.04$~meV. $\Gamma_{1S}$ and $\Gamma_{2S}$ were both tuned to be $\sim$2--3\% of $\Gamma_{1D}$ and $\Gamma_{2D}$, respectively, such that the biased leads probe the equilibrium local density of states on their respective dot. The bias is applied to both dots simultaneously. The parameters used for the calculations were the same as Fig. 1c and 1d, except with $T = 40$~mK and $B = 1$ T. Note that $\alpha_1=\alpha_2=1$ serve only as normalization factors in the calculation. The $\epsilon_1, \epsilon_2$ described an $N_e = 1$ LBTP and a particle-hole transformation was performed to describe this $N_e = 3$ LBTP (supplemental info).
\label{fig4}
}
\end{figure}
\fi

The positions of the peaks may be qualitatively understood by considering inelastic transitions between Zeeman-split states (Fig. 4a). In the Kondo effect, a bias voltage can compensate for a broken degeneracy. Given that we have broken the spin and pseudospin degeneracies, naively we expect that Kondo effect related fluctuations (anomalies) could appear at six bias voltages: $\omega = \pm E_{PZ}$, $\omega = \pm E_Z$, and $\omega = \pm (E_Z+E_{PZ})$. However, because dot 2 is fully occupied, only processes corresponding to incoming holes, at negative energy, should manifest as peaks in $G_2$ and $A_2$. $G_2$ exhibits two broad (merged) peaks positioned at $\omega = -E_{PZ}, -(E_Z+E_{PZ})$, with $A_2$ in qualitative agreement. Analogously, $G_1$ and $A_1$ exhibit a peak at negative energy, near $\omega = -E_Z$, because of the occupied state of dot 1. Dot 1 also has an empty state, so it should yield processes involving incoming electrons. Inelastic transitions involving this empty state and the three other filled states of the double dot should therefore appear as peaks at $\omega = E_{PZ}, E_Z, E_Z+E_{PZ}$. For small $E_{PZ}$, the peak at $E_{PZ}$ is visible in both $G_1$ and $A_1$. We interpret this peak as being merged with the other two expected peaks, with the peak at smallest $\omega$ dominating, such that a peak should be seen at $\omega = \textrm{min}(E_{PZ},E_Z)$ in both $G_1$ and $A_1$.

Some of these peaks are related to more familiar Kondo phenomenology. The peak at $\omega = 0$ in $G_1$ and $G_2$ for $E_{PZ} = 0$ is the ``zero-bias anomaly'' of a purely orbital Kondo state, since spin degeneracy has been broken by $E_Z$. The Zeeman-split levels appear as a peak at $\omega = -E_Z$ in $G_1$ (appearing at both $\pm E_Z$ for $E_{PZ}\ge E_Z$). As the (1,2) configuration becomes favored with increasing $E_{PZ}$, dot 1 hosts the unpaired spin. This unpaired spin gives rise to a spin Kondo resonance, reminiscent of the spin-1/2 Kondo effect, where the Kondo resonance splits into peaks at $\pm E_Z$ when a magnetic field $g\mu_{B}B > T_K$ is applied \cite{Meir1993, Cronenwett1998, Kogan2004}. For $E_{PZ} \ge E_Z$, transport through the orbital favoring the unpaired spin will exhibit the Zeeman splitting, and transport through the other orbital will exhibit the pseudo-Zeeman splitting.

The remaining mystery in the experimental data is the saturation of the conductance at $T \lesssim 40$~mK, as observed in Fig. 2d and Fig. 3. We have calibrated the base electron temperature of the leads $T_e$ based on Coulomb blockade thermometry using the same device measured here, during the same cooldown in which the data presented in Figs. 2 and 3 were taken. With only a single dot formed, we find $T_e = 22$~mK. We speculate that high frequency charge noise in this device, giving rise to fluctuations in detuning, decoheres the Kondo effect and causes an apparent saturation at $T>T_e$. Coherent oscillations of a DQD charge qubit, considered as a two-level system, happen with frequency $\Omega = \sqrt{t^2+\delta^2}/\hbar$, where $t$ is the interdot tunnel coupling and $\delta$ is the detuning. By measuring series conductance G$_{\textrm{series}}$ $<$ 0.001 e$^2$/h between the dots at the triple points, we can establish the bound $|t|<0.3$ $\mu$eV (details in \cite{Amasha2013}). We consider $t$ to be negligible, giving $\Omega = \delta/\hbar$ and a typical ``dephasing rate'' $\Gamma_\delta = \sqrt{\langle \delta^2 \rangle}/\hbar \sim 1/T_{2^*}$, where $\sqrt{\langle \delta^2 \rangle}$ is the size of the detuning fluctuations. This loss of phase coherence should result in an abrupt saturation of conductance at temperature $T \sim \sqrt{\langle \delta^2 \rangle}/ k_B$ due to a renormalization cutoff. With our experimental setup and device, we cannot directly measure our fluctuations in detuning. However, other researchers have used microwave-induced charge state repopulation to extract $\sqrt{\langle \delta^2 \rangle}$ as high as 3 $\mu$eV (=35~mK) \cite{Hayashi2003} and 3.7 $\mu$eV (=43~mK) \cite{Petersson2010} for DQDs in GaAs/AlGaAs heterostructures. This roughly corresponds to the temperature at which we observe the apparent saturation.

In conclusion, we report on the SU(4) Kondo effect in a GaAs/AlGaAs double quantum dot. We first show the importance of both spin and orbital degrees of freedom by demonstrating the particle-hole symmetry of a four-fold degenerate state. We proceed to demonstrate the exceptional agreement of experiment and theory at a LBTP, and show the expected universal SU(4) scaling. Finally, we use the pseudospin resolution afforded by this system to demonstrate how the Kondo resonance splits when the four-fold degeneracy is broken: one dot exhibits a Zeeman splitting and the other a pseudo-Zeeman splitting. These results highlight the remarkable power of using lateral quantum dots to realize and investigate quantum impurity problems.

We are grateful to Y. Oreg, A. Carmi, J. K\"{o}nig, Ali G. Moghaddam, G. B. Martins, C. A. B\"{u}sser, A. E. Feiguin, and L. Peeters for discussions. Experimental work was supported by the NSF under DMR-0906062, by the U.S.-Israel BSF grant No. 2008149, and most recently by the Gordon and Betty Moore Foundation through Grant GBMF3429. A.~J.~K. acknowledges a Stanford Graduate Fellowship. G.~Z. and C.~P.~M. acknowledge support from Hungarian Grant Nos. K105149 and CNK80991. C.~P.~M. was financially supported by UEFISCDI under French-Romanian Grant DYMESYS, Contract No. PN-II-ID-JRP-2011-1. I.~W. acknowledges support from EU grant No. CIG-303 689 and MSHE grant No. IP2011~059471. NRG calculations were performed at Pozna\'{n} Supercomputing and Networking Center.

{\bf Author contributions: }

AJK, SA, GZ, and DGG designed the experiment. AJK and SA performed the measurements, with substantial contributions from IGR. IW, CPM, and GZ performed the NRG calculations. CPM and IW contributed equally to the theoretical analysis. AJK, SA, CPM, IW, GZ, and DGG analyzed the data. SA designed and fabricated the devices, with e-beam lithography done by JK, using heterostructures grown by HS. AJK wrote the paper with critical review provided by all other authors.

{\bf Competing financial interests: }

The authors declare no competing financial interests.

\newpage

\ifarxiv
\else

\begin{figure}
\ifarxiv
\includegraphics[width=6in]{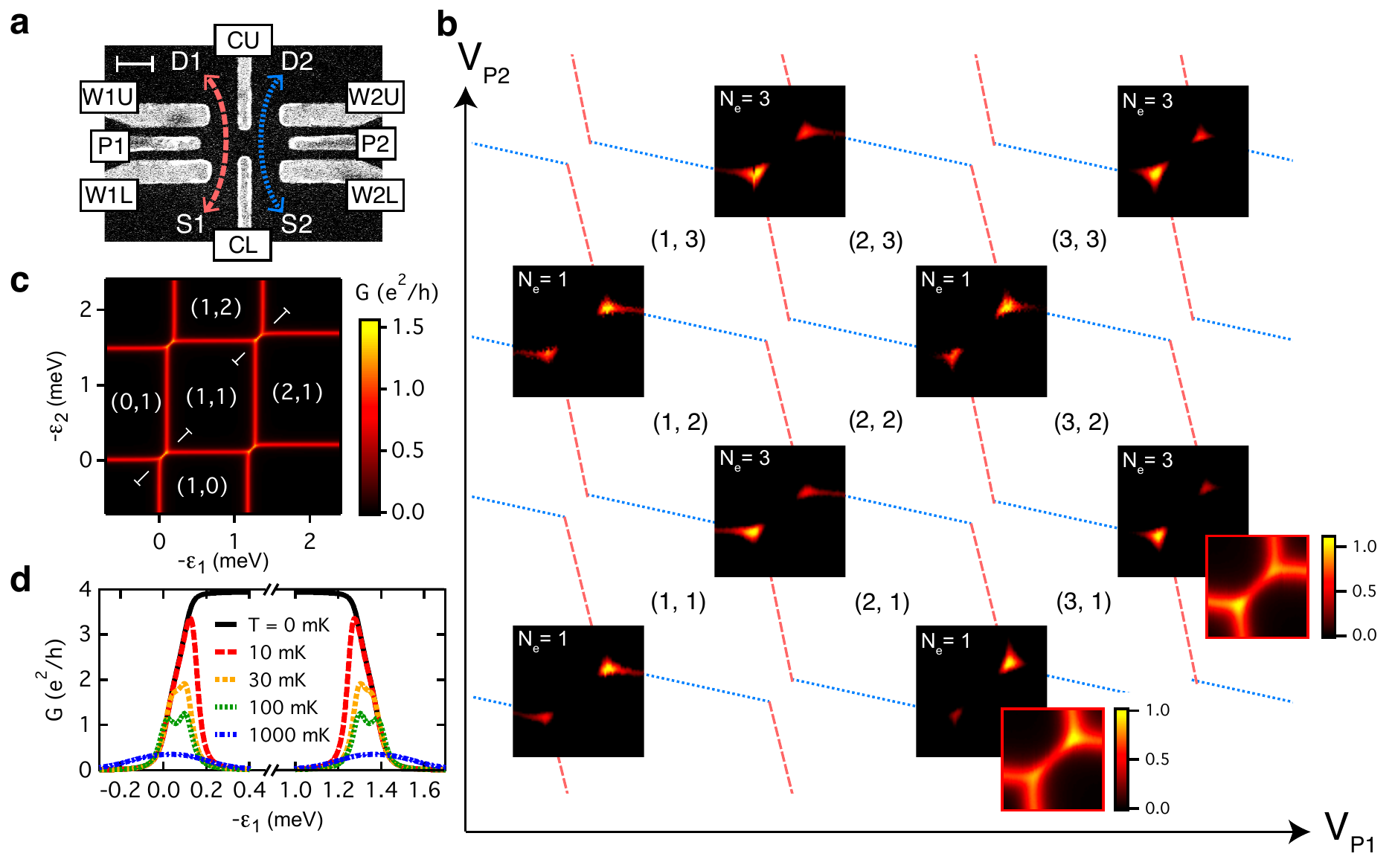}
\fi
\caption{\small {\bf Survey of conductance at orbital degeneracies.}
{\bf (a)} SEM micrograph of a double quantum dot like the one measured.  The scale bar (upper left) is 200nm. Currents are measured from S1 to D1 (red dashed arrows) and from S2 to D2 (blue dotted arrows).
{\bf (b)} Experimental conductance $G$ = $G_1$ + $G_2$ at $N_e = 1$ and $N_e = 3$ LBTPs. Electron occupancies are labeled within each hexagon, relative to a (0,0) hexagon where each dot's occupancy is only known modulo 2. Each square of data corresponds to a region spanning 3 mV in $V_{P1}$ and $V_{P2}$. The color scales are individually set so that only data between 75\%--100\% of the maximum conductance are visible. Two examples of the data with the color scale unsaturated are also shown (bottom-right). The red lines and blue lines behind the data schematically depict the charge stability diagram (not to scale). Red dashed (blue dotted) lines indicate Coulomb blockade peaks for dot 1 (2). 
{\bf (c)} Charge stability diagram from NRG calculations of $G$ = $G_1$ + $G_2$. The parameters used were $T = 30$~mK, $B =$~0, $U_1 = 1.2$~meV, $U_2 = 1.5$~meV, $U^\prime = 0.1$~meV, $\Delta_1 = 0.017$~meV, $\Delta_2 = 0.0148$~meV, $\alpha_1 = \alpha_2 = 1$.  $\epsilon_{1(2)}$ is the energy of the dot 1(2) state above the Fermi energies of the leads. The white text labels the number of electrons on dots 1 and 2. The white brackets mark the location of $N_e = 1$ (lower-left) and $N_e = 3$ (upper-right) LBTPs.
{\bf (d)} Cuts along the $N_e = 1$ and $N_e = 3$ LBTPs of (c) at zero and finite temperatures, parameterized by $-\epsilon_1$. The left part of (d) corresponds to $N_e = 1$ and the right part corresponds to $N_e = 3$. At 30 mK, the conductance is higher at the peaks nearest to the (1,1) hexagon, as in the experimental data of (b).
\label{fig1}
}
\end{figure}

\newpage
\begin{figure}
\ifarxiv
\includegraphics[width=6in]{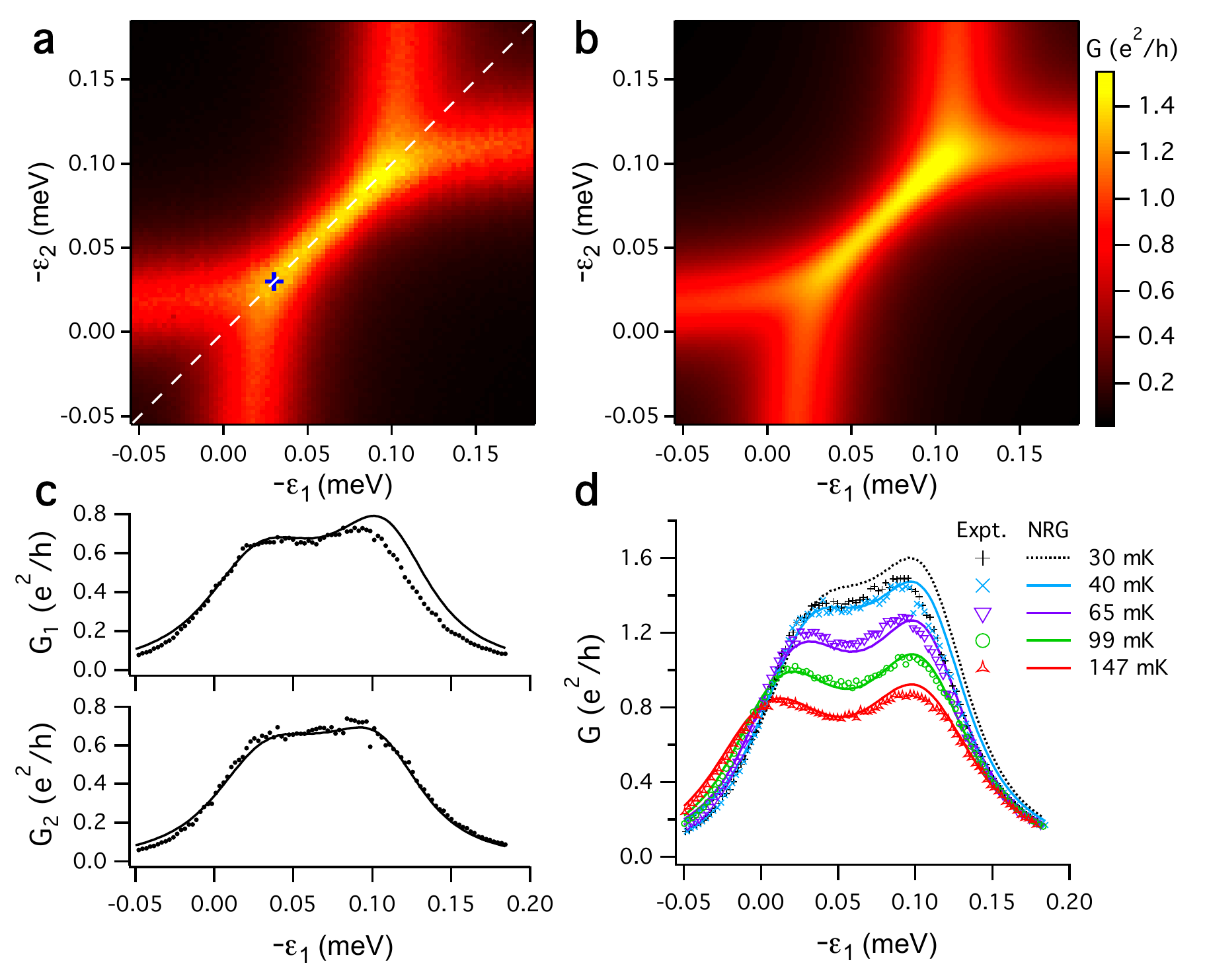}
\fi
\caption{\small {\bf Agreement between experimental data and NRG calculations.}
Experimentally measured {\bf (a)} and theoretically computed {\bf (b)} total conductance at an $N_e=1$ LBTP in units of $e^2/h$ at $T = 40$~mK. The white dashed line corresponds to zero detuning. The blue crosses mark the point $\epsilon_1 = \epsilon_2 = -0.03$~meV, where the SU(4) scaling will be demonstrated in Fig. 3. The parameters used in the NRG calculations are the same as in Fig. 1c and 1d, except with $\alpha_1 = \alpha_2 = 0.875$. 
{\bf (c)} Conductances $G_1$ (top) and $G_2$ (bottom) along the LBTP, indicated by the dashed lines in panels (a) and (b). Dots denote the experimental data, solid lines are the NRG results.  
{\bf (d)} Total conductance along the LBTP for a subset of measured temperatures. Symbols denote the experimental data, solid lines are the NRG results. The agreement between the experiment and theory over the range of energies shown is typical for $T \ge 40$~mK, but the experimental data saturate below 40~mK.
\label{fig2}
}
\end{figure}
\newpage
\begin{figure}
\ifarxiv
\includegraphics[width=6in]{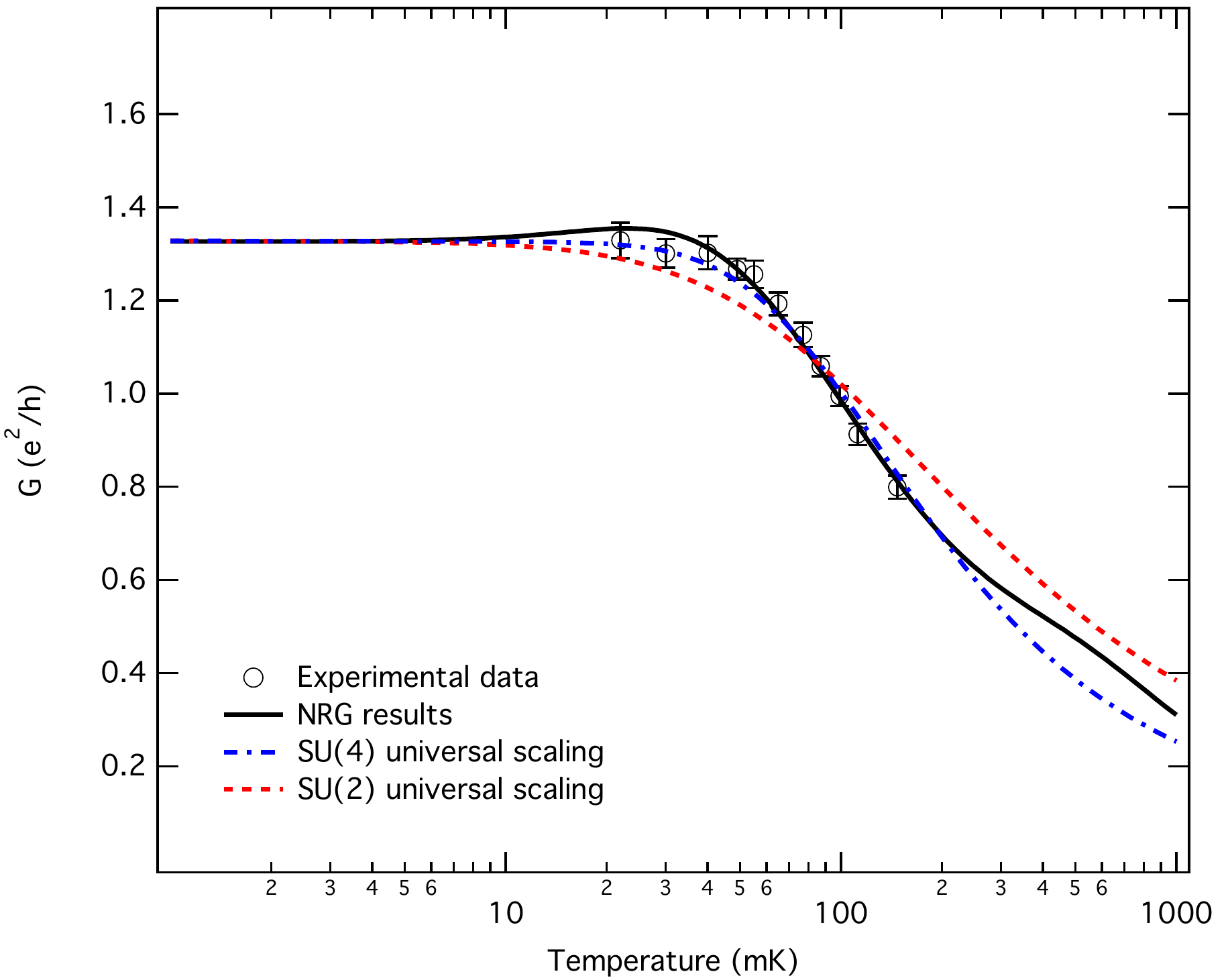}
\fi
\caption{\small {\bf Universal scaling of the conductance.}
Experimental data for the temperature dependence of the conductance (symbols) 
at $\epsilon_1 = \epsilon_2 = -0.03$~meV in Fig. 2d. Experimental data are compared with NRG results as well as with the universal SU(4) and SU(2) curves using best-fit Kondo temperatures $T_{K SU(2)} = 320$~mK and $T_{K SU(4)} = 220$~mK.
\label{fig3}
}
\end{figure}
\newpage
\begin{figure}
\ifarxiv
\includegraphics[width=6in]{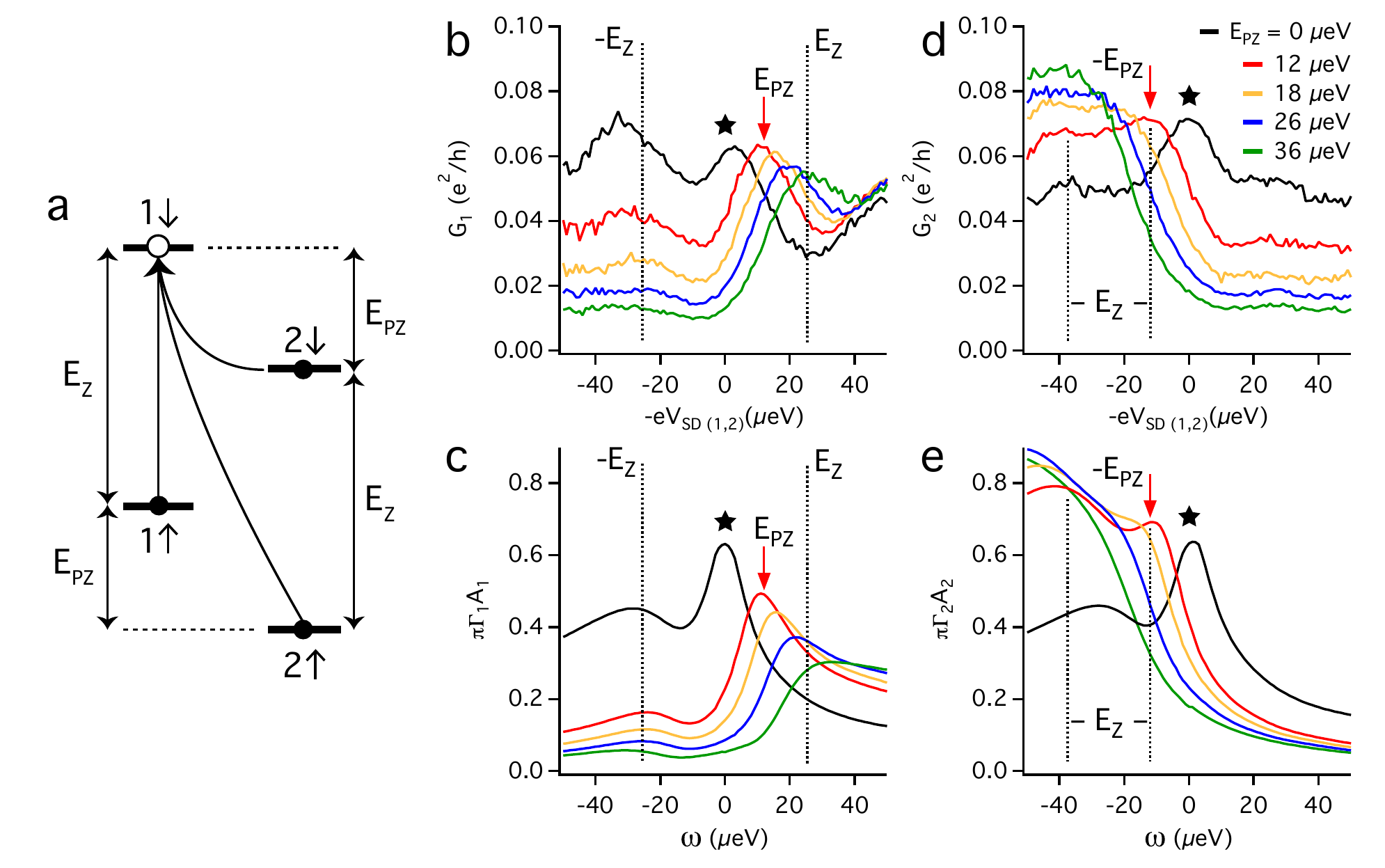}
\fi
\caption{\small {\bf Orbital state-resolved bias spectroscopy of the SU(4) Kondo resonance.}
{\bf (a)} Inelastic transitions between Zeeman-split states of dot 1 and dot 2 at an $N_e = 3$ LBTP.
{\bf (b)} Experimental conductance $G_1$ for dot 1 in a 1.0 T Zeeman field.  The five traces correspond to different values of $E_{PZ}$, with $E_{PZ}>0$ meaning dot 1 is favored to hold the unpaired electron. A predicted ``spin-Kondo like'' feature at $-eV_{SD} \approx -E_Z$ is observed. The star marks a purely orbital Kondo state for $E_{PZ} = 0$ which appears in both dots.
{\bf (c)} Calculated spectral function $A_1$ for dot 1 in a 1.0 T Zeeman field. 
{\bf (d)} Experimental conductance $G_2$ for dot 2 in a 1.0 T Zeeman field. For the same increments in $E_{PZ}$, no peaks are pinned to a particular energy, and instead they move with the pseudo-Zeeman splitting.
{\bf (e)} Calculated spectral function $A_2$ for dot 2 in a 1.0 T Zeeman field. 
For all panels, $\Gamma_1, \Gamma_2 \approx 0.04$~meV. $\Gamma_{1S}$ and $\Gamma_{2S}$ were both tuned to be $\sim$2--3\% of $\Gamma_{1D}$ and $\Gamma_{2D}$, respectively, such that the biased leads probe the equilibrium local density of states on their respective dot. The bias is applied to both dots simultaneously. The parameters used for the calculations were the same as Fig. 1c and 1d, except with $T = 40$~mK and $B = 1$ T. Note that $\alpha_1=\alpha_2=1$ serve only as normalization factors in the calculation. The $\epsilon_1, \epsilon_2$ described an $N_e = 1$ LBTP and a particle-hole transformation was performed to describe this $N_e = 3$ LBTP (supplemental info).
\label{fig4}
}
\end{figure}

\fi


\begin{thebibliography}{99}
\bibitem{Note} Though the bare exchange couplings are not SU(4) symmetrical, at low temperatures the effective exchange coupling is renormalized and becomes more and more SU(4) symmetrical.
\bibitem{footnote1} In contrast, $G=G_1+G_2$ is much less sensitive to the cut direction.
\end{thebibliography}
\end{document}


\newcommand{\bra}[1]{\langle #1|}
\newcommand{\ket}[1]{|#1\rangle}
\newcommand{\braket}[2]{\langle #1|#2\rangle}
\newcommand{\brag}{\langle\langle}
\newcommand{\ketg}{\rangle\rangle}
\newcommand{\s}{\sigma}
\newcommand{\clebsch}[3]{\langle #1;#2 \vert #3\rangle}
\newcommand{\expect}[1]{\langle #1 \rangle}
\newcommand{\e}{\varepsilon}
\newcommand{\w}{\omega}
\newcommand{\G}{\Gamma}
\newcommand{\up}{\uparrow}
\newcommand{\rt}{\rightarrow}
\newcommand{\down}{\downarrow}

\newcommand{\beq}{\begin{eqnarray}}
\newcommand{\eeq}{\end{eqnarray}}
\newcommand{\be}{\begin{equation}}
\newcommand{\ee}{\end{equation}}
\newcommand{\de}{{\rm d }}
\newcommand{\dd}{\partial}
\newcommand{\h}{\hslash}

\renewcommand{\thefigure}{S\arabic{figure}}
\renewcommand{\thetable}{S\arabic{table}}
\renewcommand{\thesection}{S\arabic{section}}

\title{\large \bf Supplemental information for  \linebreak``Observation of the SU(4) Kondo state in a \linebreak double quantum dot''}
\author[1]{\normalsize A.~J.~Keller}
\author[1,$\dag$]{S.~Amasha}
\author[2]{I.~Weymann}
\author[3,4]{C.~P.~Moca}
\author[1,$\ddag$]{I.~G.~Rau}
\author[5]{J.~A.~Katine}
\author[6]{Hadas~Shtrikman}
\author[3]{G.~Zar\'{a}nd}
\author[1,*]{D.~Goldhaber-Gordon}
\affil[1]{\footnotesize Geballe Laboratory for Advanced Materials, Stanford University, Stanford, CA 94305, USA}
\affil[2]{Faculty of Physics, Adam Mickiewicz University, Pozna\'n, Poland}
\affil[3]{BME-MTA Exotic Quantum Phases ``Lend\"{u}let'' Group, Institute of Physics, Budapest University of Technology and Economics, H-1521 Budapest, Hungary}
\affil[4]{Department of Physics, University of Oradea, 410087, Romania}
\affil[5]{HGST, San Jose, CA 95135, USA}
\affil[6]{Department of Condensed Matter Physics, Weizmann Institute of Science, Rehovot 96100, Israel}
\affil[$\dag$]{Present address: MIT Lincoln Laboratory, Lexington, MA 02420, USA}
\affil[$\ddag$]{Present address: IBM Research -- Almaden, San Jose, CA 95120, USA}
\affil[*]{Corresponding author; goldhaber-gordon@stanford.edu}
\date{}
\maketitle
\tableofcontents
\newpage

\section{Full LBTP survey}

\begin{landscape}
\begin{figure}
\begin{center}
\includegraphics[width=6.75in]{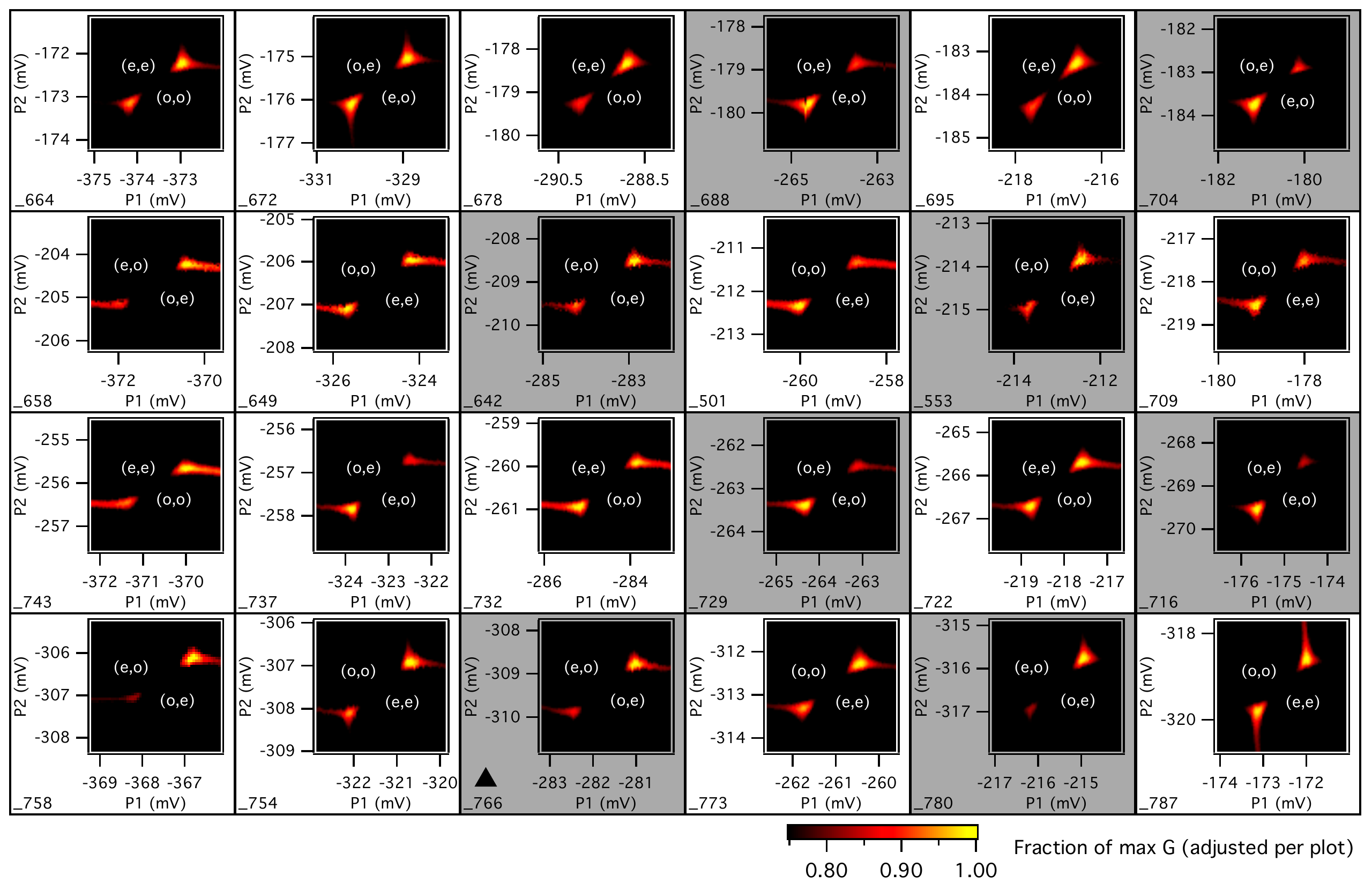}
\caption{\small Experimental zero bias conductance $G = G_1 + G_2$ for a survey of 24 LBTPs, at zero magnetic field and at T = 20 mK. Electron occupation numbers are labeled here by their parity (e = even, o = odd). Each square of measured data corresponds to a region spanning 3 mV in $V_{P1}$ and $V_{P2}$. The color scales for each square have been individually set so that only data between 75\% and 100\% of the maximum conductance are visible. Each data set is identified by a number in the bottom-left of each plot. Set 766 (marked by triangle) corresponds to the bottom-left plot in Fig. 1d. All sets appearing in Fig. 1d have a gray background.  Of the four other odd parity data sets, only one (672) does not show the expected asymmetry; it has no clear asymmetry at all.\label{surveyB0}}
\end{center}
\end{figure}
\end{landscape}

The data presented in Fig. 1b are only a subset of the full survey of conductance around lines between triple points (LBTPs). The full survey, shown in Fig. \ref{surveyB0}, demonstrates that 11/12 of $N_e = 1$ or $N_e = 3$ LBTPs exhibit higher conductance towards the adjacent (1,1) hexagon. In addition, twelve (1,1)/(2,0) or (0,2)/(1,1) LBTPs were surveyed: these should possess a five-fold degeneracy assuming the (2,0) ground state is a singlet rather than triplet. The $N_e = 1$ and $N_e = 3$ LBTPs differ qualitatively from the (1,1)/(2,0) and (0,2)/(1,1) LBTPs in that the latter class of LBTPs do not exhibit a simple pattern of which end of the LBTP has higher conductance. Experimental parameters $\Gamma_1$, $\Gamma_2$ and peak conductances are extracted from each data set and summarized in Table \ref{surveyb0tab}.

\begin{landscape}
\begin{figure}
\begin{center}
\includegraphics[width=6.75in]{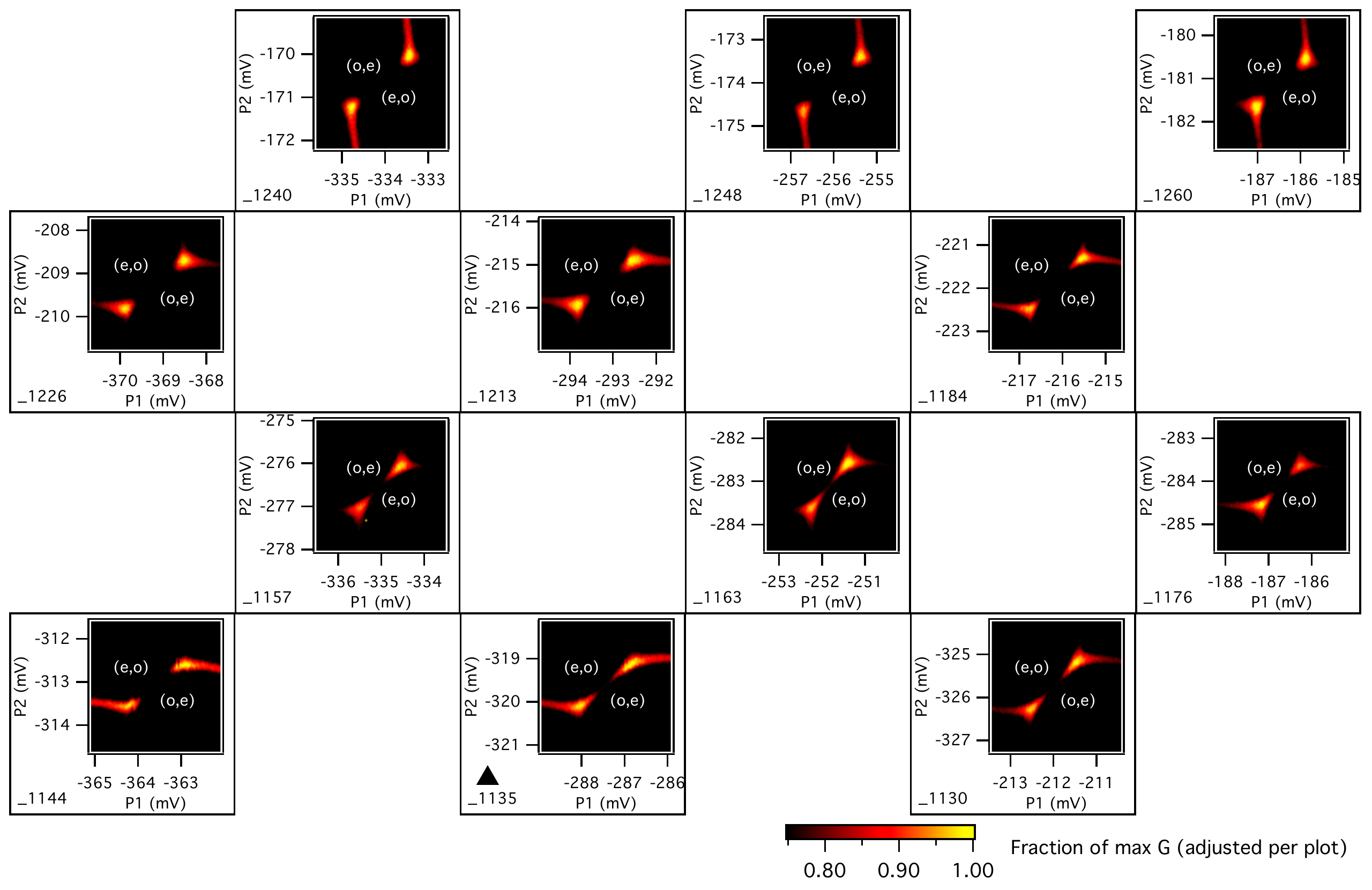}
\caption{Experimental zero source-drain bias conductance $G = G_1 + G_2$ for a survey of twelve $N_e = 1$ and $N_e = 3$ LBTPs, in an in-plane magnetic field of 2.0 T at T = 20 mK. The LBTPs surveyed correspond to the same absolute electron occupations as the LBTPs of Fig. \ref{surveyB0}. The data are presented as described in the caption of Fig. \ref{surveyB0}. Set 1135 (marked by triangle) corresponds to the bottom-left plot in Fig. 1b, and set 766 in Fig. \ref{surveyB0}. \label{surveyB2}}
\end{center}
\end{figure}
\end{landscape}

Because we claim that the $N_e = 1$ and $N_e = 3$ LBTP data reflect the particle-hole symmetry of a four-fold degenerate state, it is natural to expect that the pattern is destroyed when the four-fold degeneracy is broken. Fig. \ref{surveyB2} shows the $N_e = 1$ and $N_e = 3$ LBTPs surveyed again in an in-plane magnetic field of 2.0 T, corresponding to $E_Z = g \mu_B B = 0.051$~meV for $g = 0.44$. Here, $E_Z > \Gamma_1$, $\Gamma_2$ for all of the surveyed LBTPs. With the Zeeman splitting having broken the spin degeneracy at the LBTPs, a periodic pattern is no longer discernible. Table \ref{surveyb2tab} summarizes the extracted parameters for each data set, as in Table \ref{surveyb0tab}.

\begin{landscape}
\begin{figure}
\begin{center}
\includegraphics[width=8.5in]{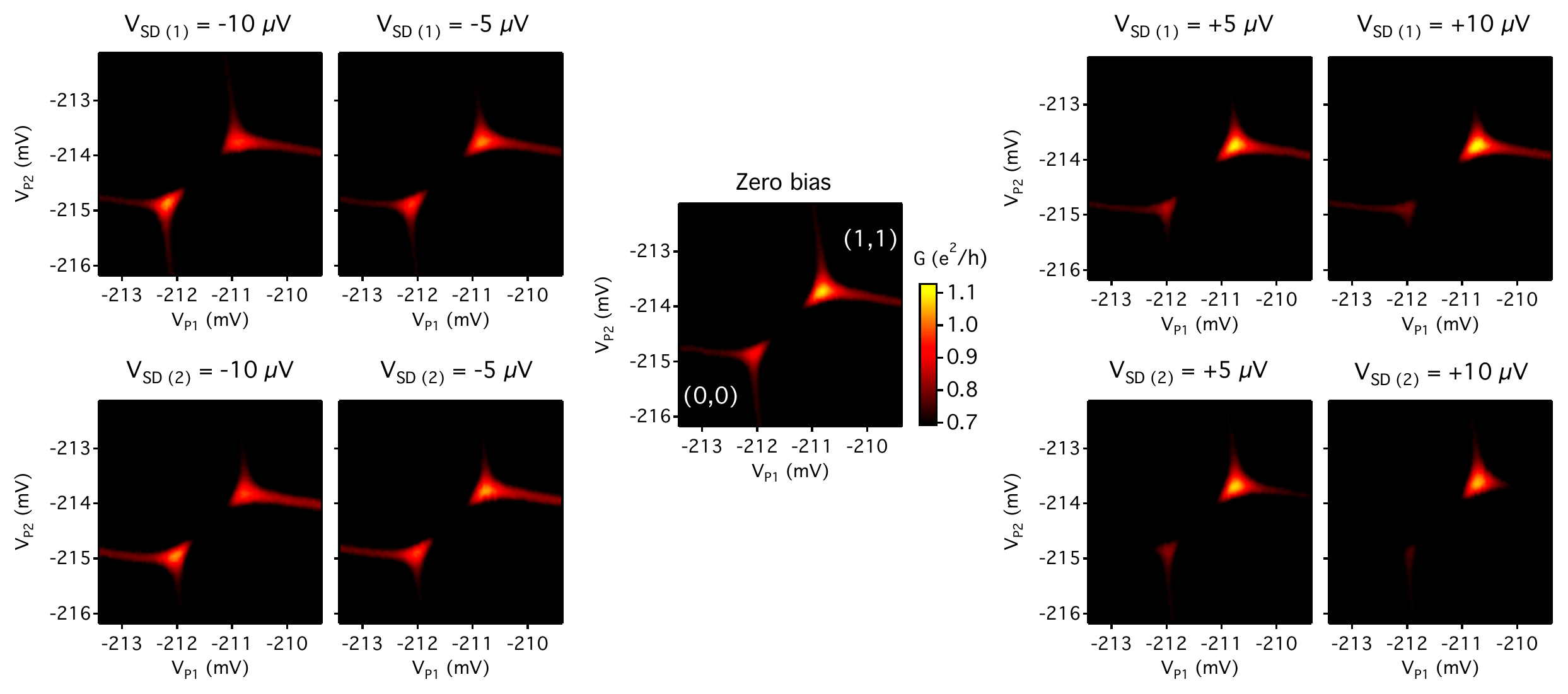}
\caption{Experimental conductance $G = G_1+G_2$ measured at an $N_e=1$ LBTP, with small but finite $V_{SD}$. All color scales show from 0.70 to 1.12 $e^2/h$, emphasizing the conductance along the LBTP near (0,0) and (1,1). Center: $V_{SD}=0$ for each dot. This data set was taken immediately after checking the input offset voltage of both current amplifiers. Top row: Finite $V_{SD}$ is applied across dot 1 only. Bottom row: Finite $V_{SD}$ is applied across dot 2 only. 
\label{biasdep}}
\end{center}
\end{figure}
\end{landscape}

Fig. \ref{biasdep} shows how a small but finite $V_{SD}$ affects the observed asymmetry at an $N_e = 1$ LBTP. The LBTP measured here corresponds to the same absolute electron occupation numbers as data set 553 shown in Fig. \ref{surveyB0}. Only for negative $V_{SD}$ approaching $-10$ $\mu$V does the conductance near (0,0) exceed that near (1,1). For positive $V_{SD}$, the pattern of higher conductance nearer to (1,1) than (0,0) is actually exaggerated. The effect of finite $V_{SD}$ is similar regardless of whether it is applied to dot 1 or 2. Input offset voltages from current amplifiers could obscure our observed pattern, were it not for our ability to stabilize these voltages to within 1 $\mu$V (see section \ref{electronics}). 

\begin{table}
\begin{center}
\begin{tabular}{|c|c|c|c|c||c|c|c|c|c|}
\hline
Data set & $\Gamma_1$ & $\Gamma_2$ & $\gamma_1$ & $\gamma_2$ & Data set & $\Gamma_1$ & $\Gamma_2$ & $\gamma_1$ & $\gamma_2$ \\
\hline
\_664 & 24 & 32 & 0.82 & 0.89 & \_743 & 27 & 31 & 0.43 & 0.80 \\
\_672 & 29 & 36 & 0.69 & 0.66 & \_737 & 27 & 29 & 0.66 & 0.77 \\
\_678 & 35 & 39 & 0.63 & 0.68 & \_732 & 27 & 31 & 0.56 & 0.78 \\
\_688 & 28 & 36 & 0.54 & 0.67 & \_729 & 28 & 31 & 0.59 & 0.79 \\
\_695 & 32 & 42 & 0.67 & 0.70 & \_722 & 30 & 33 & 0.69 & 0.80 \\
\_704 & 33 & 32 & 0.73 & 0.78 & \_716 & 32 & 34 & 0.74 & 0.80 \\
\_658 & 26 & 27 & 0.51 & 0.89 & \_758 & 28 & 33 & 0.48 & 0.68 \\
\_649 & 30 & 27 & 0.66 & 0.88 & \_754 & 30 & 31 & 0.64 & 0.67 \\
\_642 & 28 & 27 & 0.75 & 0.90 & \_766 & 27 & 30 & 0.51 & 0.64 \\
\_501 & 26 & 31 & 0.51 & 0.84 & \_773 & 27 & 34 & 0.59 & 0.68 \\
\_553 & 28 & 29 & 0.82 & 0.89 & \_780 & 29 & 34 & 0.73 & 0.67 \\
\_709 & 30 & 33 & 0.74 & 0.92 & \_787 & 29 & 34 & 0.77 & 0.66 \\
\hline
\end{tabular}
\caption{For each data set shown in Fig. \ref{surveyB0}, experimentally controllable parameters $\Gamma_1$, $\Gamma_2$, $\gamma_1$, and $\gamma_2$ are extracted by fitting a Lorentzian lineshape to a Coulomb blockade (CB) peak neighboring the LBTP. $\Gamma_{1(2)}$ corresponds to the FWHM of the CB peak in dot 1 (2), in units of $\mu$eV. The width in gate voltage is converted to an energy using conversion factors derived from bias spectroscopy, taken near each LBTP. $\gamma_{1(2)}$ are defined to equal the conductance at the CB peak of dot 1 (2) in $e^2/h$. For these data it is not known whether the source or drain lead is more coupled for either dot. In all cases, the electron temperature $T_e = 20$~mK.
\label{surveyb0tab}}
\end{center}
\end{table}

\begin{table}
\begin{center}
\begin{tabular}{|c|c|c|c|c|}
\hline
Data set & $\Gamma_1$ & $\Gamma_2$ & $\gamma_1$ & $\gamma_2$ \\
\hline
\_1240 & 29 & 33 & 0.98 & 0.70\\
\_1248 & 29 & 34 & 0.88 & 0.65\\
\_1260 & 31 & 36 & 0.90 & 0.80\\
\_1226 & 32 & 32 & 0.62 & 0.78\\
\_1213 & 30 & 35 & 0.70 & 0.83\\
\_1184 & 31 & 32 & 0.76 & 0.87\\
\_1157 & 34 & 32 & 0.94 & 0.95\\
\_1163 & 31 & 35 & 0.94 & 0.99\\
\_1176 & 35 & 31 & 0.88 & 0.98\\
\_1144 & 32 & 29 & 0.58 & 1.02\\
\_1135 & 32 & 31 & 0.75 & 0.99\\
\_1130 & 30 & 35 & 0.79 & 0.97\\
\hline
\end{tabular}
\caption{For each data set shown in Fig. \ref{surveyB2}, experimentally controllable parameters $\Gamma_1$, $\Gamma_2$, $\gamma_1$, and $\gamma_2$ are extracted and reported as in Table \ref{surveyb0tab}. \label{surveyb2tab}}
\end{center}
\end{table}

\section{Summary of NRG calculations}

\subsection{NRG calculations}

In our numerical calculations the double quantum dot (DQD) system is modeled by
the following Hamiltonian
\be
H = H_{\rm DQD} + H_{\rm Tun} + H_{\rm Leads},
\ee
where 
\beq
  H_{\rm DQD} &=& \sum_{j\s} \e_{j} n_{j\s} + \sum_j U_j n_{j\up}n_{j\down} \nonumber\\
  &+& U'\sum_{\s\s'} n_{1\s} n_{2\s'} + g\mu_B B_z S_z ,
\eeq
describes the two dots, with $n_{j\s} = d_{j\s}^\dag d_{j\s}$ the occupation
number operator of dot $j=1,2$ for spin $\s$, $\e_{j\s}$ the energy
of a spin-$\s$ electron residing on dot $j$. $U_j$ ($U'$) denotes
the intradot (interdot) Coulomb correlations, while $B_z$ is the magnetic field
applied along the $z$-direction and $S_z$ is the $z$-component of the double dot's spin.
The tunneling Hamiltonian $H_{\rm Tun}$ reads
\be
H_{\rm Tun} = \sum_{\alpha k} \sum_{j\s} t_{\alpha j} (c_{\alpha jk\s}^\dag d_{j\s} + d_{j\s}^\dag c_{\alpha jk\s}),
\ee
where $c_{\alpha j k\s}^\dag$ is the creation operator of an electron
in lead $\alpha = L,R$ coupled to dot $j$,
with momentum $k$ and spin $\s$ of energy $\e_{\alpha j k}$.
Tunneling processes between the dots and leads are described
by hopping matrix elements $t_{\alpha j}$. Tunneling between the two dots is suppressed by tuning gates in our experiment, and hence is omitted from the model.
The leads are described by noninteracting quasiparticles
\be
H_{\rm Leads} = \sum_{\alpha j k\s} \e_{\alpha j k} c_{\alpha j k\s}^\dag c_{\alpha j k\s} .
\ee
Due to the coupling to external leads, the dots' levels acquire a width
described by $\Delta_{\alpha j} = \pi \rho_{\alpha j} |t_{\alpha j}|^2$,
with $\rho_{\alpha j}$ the density of states of lead $\alpha$ coupled to dot $j$.

We performed the full density-matrix numerical renormalization group calculations (fDM-NRG)
\cite{WilsonRMP75,BullaRMP08,WeichselbaumPRL07,TothPRB08},
employing the Budapest Flexible DM-NRG code~\cite{FlexibleDMNRG}.
For efficient calculations, we used the charge $U(1)$ and the spin $SU(2)$ symmetries in each channel,
resulting in four symmetries altogether. When considering the effect of external magnetic field $B_z$,
the spin invariance is reduced to the $U(1)$ symmetry for the spin $z$-component in each channel.
In our computations we retained $2500-5000$ states at each iteration depending on the exploited symmetries
and used the discretization parameter $\Lambda=2$.

We calculated the linear conductance through dot $j$
using the following formula
\be
  G_j = \frac{e^2}{h}
 \alpha_j \Delta_j
  \sum_\s\int \!\!
  d\omega \; \pi A_{j\s}(\omega)\left(-\frac{\partial
  f(\omega)}{\partial\omega}\right),
\ee
where $f(\omega)$ is the Fermi-Dirac distribution function
and $\alpha_j = 4\Delta_{Lj}\Delta_{Rj}/ (\Delta_{Lj}+\Delta_{Rj})^2$ is the left-right asymmetry factor for dot $j$,
with $\Delta_j =  \Delta_{Lj} + \Delta_{Rj}$.
$A_{j\s}(\omega)$ denotes the spectral function of the $j$-th dot level
for spin $\s$, $A_{j\s}(\omega) = -\frac{1}{\pi}{\rm Im}G^R_{j\s}(\omega)$, with
$G_{j\s}^R(\omega)$ the Fourier transform of the 
retarded Green's function, $G_{j\s}^R(t)= -i\Theta(t)
\expect{\{d_{j\s}(t), d_{j\s}^\dag(0)\}}$.
To improve the quality of the spectral functions and reduce 
the effects related with broadening of Dirac delta functions,
we also used the z-averaging trick~\cite{oliveira}.

\subsection{Choosing NRG parameters}

Most of the parameters used in NRG calculations may be extracted from routine measurements of the two dots. To a good approximation, a small decrement in the dot level is proportional to a small increment in gate voltage. The proportionality constant, as well as the charging energies $U^\prime$, $U_1$, and $U_2$, are measured directly by routine bias spectroscopy.  $U^\prime$ may be extracted from the change in $\epsilon_1$ of dot 1's Coulomb blockade peak position as an electron is added to dot 2, or vice versa. $U_1$ and $U_2$ are determined from Coulomb blockade diamonds taken over a wider range of energy; results of the conductance calculations around the LBTP are largely insensitive to values of $U_1$ and $U_2$ as they are much greater than $U^\prime$.

$\Delta_1$ and $\Delta_2$ define the coupling strength (or linewidth) for dot 1 and 2 in an underlying Anderson model. $\Delta_1$ may be extracted by taking cuts away from the LBTP on a mixed valence peak of dot 1 (side of charge stability hexagon). There, for large intradot interactions $U_1$ and $U_2$, the FWHM of the conductance curve $\Gamma_1$, divided by T, must be a universal function of $\Delta_1/T$, and likewise for dot 2. In principle, for an experimentally measured $\Gamma$ at known temperature $T$, $\Delta$ should be specified by NRG calculations of that universal function. In practice, however, the $\Delta$ parameters may require some fine tuning of order 10\% for best agreement, as other effects may affect the widths of the measured peaks (perhaps Fano interference at zero magnetic field, or neglected internal states of the dots, etc.).

Effectively, the NRG calculations use two free parameters, the asymmetry parameters $\alpha_1$ and $\alpha_2$. These are selected such that the calculations reproduce the experimentally observed height of the mixed valence peaks of dot 1 and 2, as well as the temperature dependent conductance in other regions of parameter space. 

In Fig. 4, most of the parameters used for the spectral function calculation were unchanged from those used in NRG calculations earlier in the paper. However, in the calculation we set $\alpha_1=\alpha_2=1$ for simplicity, as it would only contribute a scale factor otherwise. For each value of $E_{PZ}$, the corresponding values of $\epsilon_1$ and $\epsilon_2$ are shown in Table \ref{epsilons}.

For computational convenience we treated an $N_e = 1$ LBTP. However, by means of a particle-hole transformation ($\omega \rt -\omega$, $d_{j\sigma}\leftrightarrow d_{j\sigma}^\dagger$, $\epsilon_1 \rt \epsilon_1 - U_1 - 2U^\prime$, $\epsilon_2 \rt \epsilon_2 - U_2 - 2U^\prime$), we use these calculations to describe the $N_e = 3$ LBTP. The spectral functions shown in Fig. 4 are the result of this particle-hole transformation. For this data set, the precise values of $\Delta_1$ and $\Delta_2$ were not determined, as the tuning of the device was different from when the data for Figs. 2 and 3 were taken. Nonetheless, the $\Delta$ values should be similar and the spectral functions describe the data remarkably well.

\section{Extracting LBTP cuts from 2D data sets}

The zero-detuning cuts presented in Fig. 2c and 2d were extracted numerically from 2D data sets. The cuts are highly sensitive to cut direction such that adjusting the endpoints by even a few $\mu$eV can result in significantly different conductances along the cut. With experimental data alone, this poses a significant problem, since the line of zero detuning cannot be exactly identified. Moreover, it is difficult to control for shifts of the LBTP unrelated to renormalization as the temperature is varied. Physically meaningful shifts of the mixed-valence peaks with temperature are to be expected, but undesirable shifts, predominantly from random charge transitions in the donor layer of the heterostructure, may also contribute.

To address these concerns, for fixed NRG parameters we compare the 2D experimental data sets to the 2D NRG calculations, at each measured temperature. The pseudospin-resolved conductances from the experimental data and from NRG were fit to Lorentzians to find the peak positions. The experimental data were then offset such that the peak positions matched those in the NRG data. Some manual shifts of 0.005~meV or less were used following the fitting procedure to provide best agreement along the LBTP cuts. Note that the scale factor between gate voltage and energy is experimentally determined, and only the offsets of the axes are adjusted.

\section{Temperature dependence details}

\begin{figure}[p]
\begin{center}
\includegraphics[width=5in]{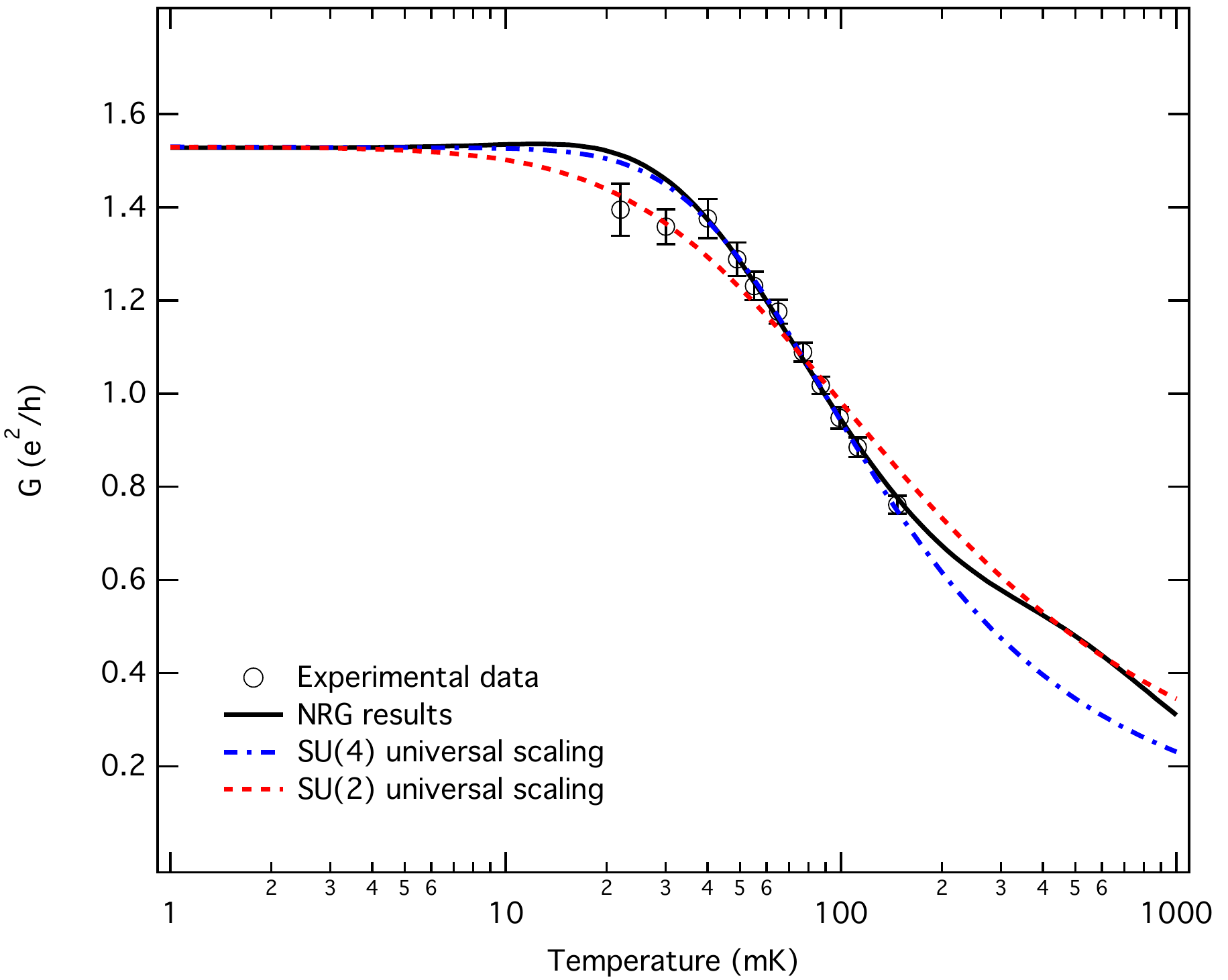}
\caption{Experimental data for the temperature dependence of the conductance (circles) at $\epsilon_1 = \epsilon_2 = -0.04$~meV in Fig. 2d. Experimental data are compared with NRG results as well as with the universal SU(4) and SU(2) curves using best-fit Kondo temperatures $T_{K SU(2)}$ = 202 mK and $T_{K SU(4)}$ =155 mK. Parameters for the NRG computations were: $B = 0$, $U_1 = 1.2$ meV, $U_2 = 1.5$ meV, $U =$ 0.1 meV, $\Delta_1 =$ 0.017 meV, $\Delta_2 =$ 0.0148 meV, $\alpha_1 = \alpha_2$ = 0.875. These are the same used in Fig. 3. \label{mev004}}
\end{center}
\end{figure}

\begin{figure}[p]
\begin{center}
\includegraphics[width=5in]{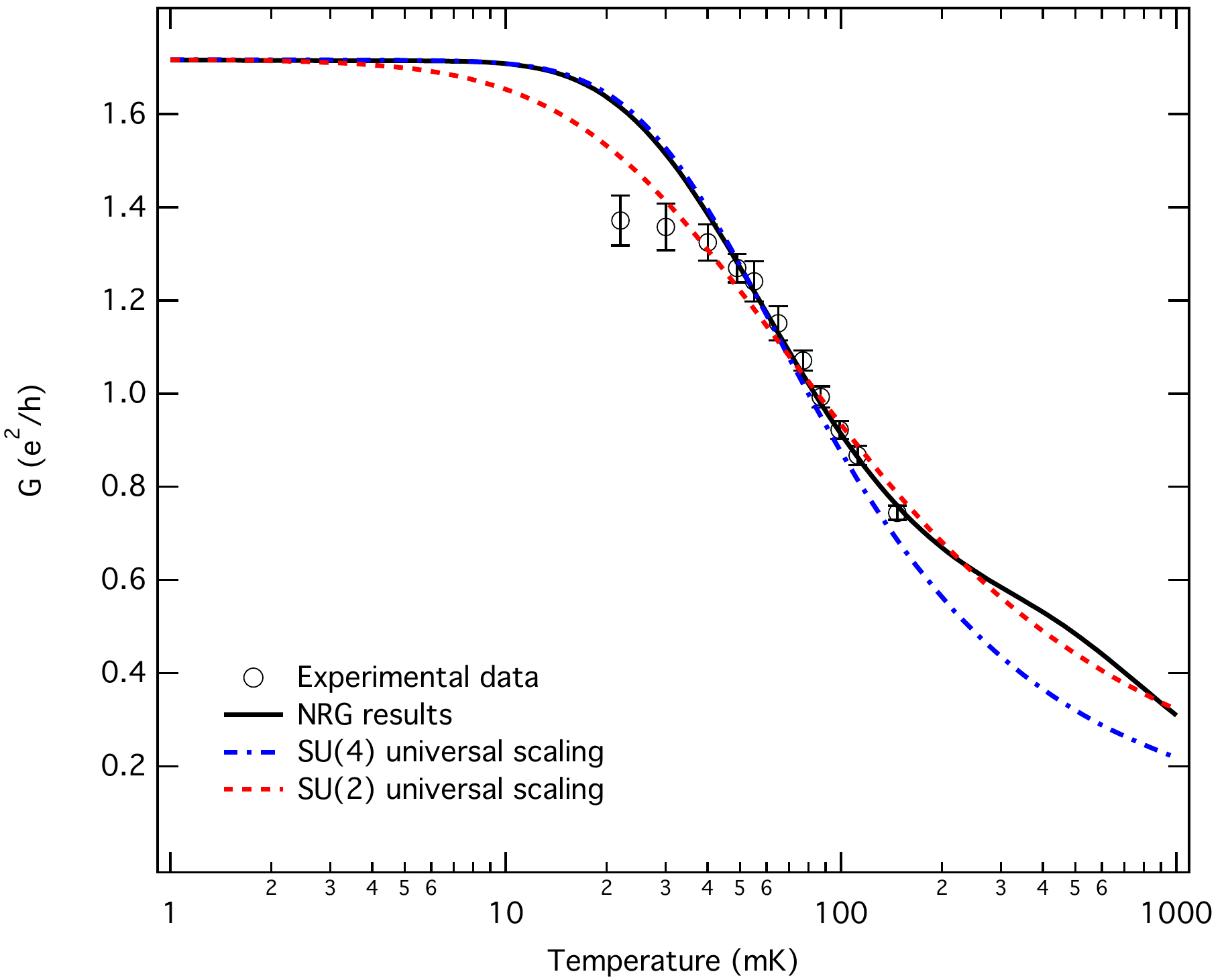}
\caption{Experimental data for the temperature dependence of the conductance (circles) at $\epsilon_1 = \epsilon_2 = -0.05$~meV in Fig. 2d. Experimental data are compared with NRG results as well as with the universal SU(4) and SU(2) curves using best-fit Kondo temperatures $T_{K SU(2)}$ = 132mK and $T_{K SU(4)}$ = 111mK. Parameters for the NRG computations were the same as in Fig. \ref{mev004}. \label{mev005}}
\end{center}
\end{figure}

As stated in the main text, the point $\epsilon$ = -0.03 meV was chosen for the temperature dependence because it is a point where $T_K$ is large compared to experimentally accessible temperatures. However, apart from the saturation observed at $T =$ 40 mK that prevents observation of the low-T rollover, the experimental data are consistent with both SU(4) universal scaling and NRG calculations for our device configuration at other points along the LBTP. In Figs. \ref{mev004} and \ref{mev005} we show the temperature dependence at $\epsilon$ = -0.04 meV and $\epsilon$ = -0.05 meV, respectively.

Uncertainties in the experimental conductances of Fig. 3 are likely dominated by the uncertainty in maintaining constant $\epsilon_1$ and $\epsilon_2$ between data taken at different temperatures, rather than conductance noise. We extract the conductances from the 2D maps of Figs. 2a and 2b and similar maps at other temperatures. The offsets (but not the scale) of the $\epsilon_1$ and $\epsilon_2$ experimental axes of Figs. 2a and 2b are set using the theoretical calculations, and this considerably reduces this uncertainty. After this alignment procedure, the remaining uncertainty in $\epsilon_1$ and $\epsilon_2$ may be conservatively taken as the pixel spacing of $\epsilon_1$ and $\epsilon_2$ in our 2D conductance maps, approximately 0.003 meV.

In determining error bars, experimental points in the 2D conductance map neighboring $\epsilon_1 = \epsilon_2 = -0.03$~meV are considered to be independent measurements of the conductance at $\epsilon_1 = \epsilon_2 = -0.03$~meV, with a Gaussian weight: $w_i = exp[-((\epsilon_1-(-0.03))^2+(\epsilon_2-(-0.03))^2)/\sigma^2]$, where $\sigma=0.003$~meV. The error bars then reflect the standard deviation of the weighted mean, and are largest at low temperatures where the conductance varies the most rapidly in any direction in $\epsilon_1$ and $\epsilon_2$. The (unbiased) standard deviation of the weighted mean, $s$, is given by:

\begin{equation}
s^2 = \frac{V_1}{V_1^2-V_2} \Sigma_{i=1}^{N} w_i (x_i - \mu^*)^2
\end{equation}

where $\mu^*$ is the weighted mean, $V_1 = \Sigma_{i=1}^N w_i$, and $V_2 = \Sigma_{i=1}^N w_i^2$.

\section{Empirical Kondo forms}
The empirical Kondo form was introduced by D. Goldhaber-Gordon, {\em et al.} \cite{goldhaber-gordon} and provides a convenient approximation of conductance through a quantum dot in the SU(2) crossover regime as a function of temperature:
\begin{equation}
G(T) = G_0 \left( 1+ (2^{1/s}-1) \left( \frac{T}{T_K} \right)^n \right)^{-s}_{\textrm{,}}
\label{su2form}
\end{equation}
where $s = 0.22$, $n=2$, $G_0$ is the conductance attained at zero temperature, and $T_K$ is the Kondo temperature. This form is purely phenomenological and was invented to describe succinctly the numerically-calculated spin-1/2 SU(2) universal scaling \cite{costi}. With such a formula it is convenient to estimate $T_K$ from experimental results using nonlinear regression, however care must be taken in its application. Importantly, for $s = 0.22$ and $n=2$ this formula does not describe the universal SU(4) scaling. Various papers have nonetheless used the empirical SU(2) form (\ref{su2form}) to fit data for which the applicability is not clear. In the absence of an alternative, this is a reasonable heuristic since the differences between the SU(4) and SU(2) scaling are subtle, but this procedure is not strictly justified.

\begin{figure}[p]
\begin{center}
\includegraphics[width=6in]{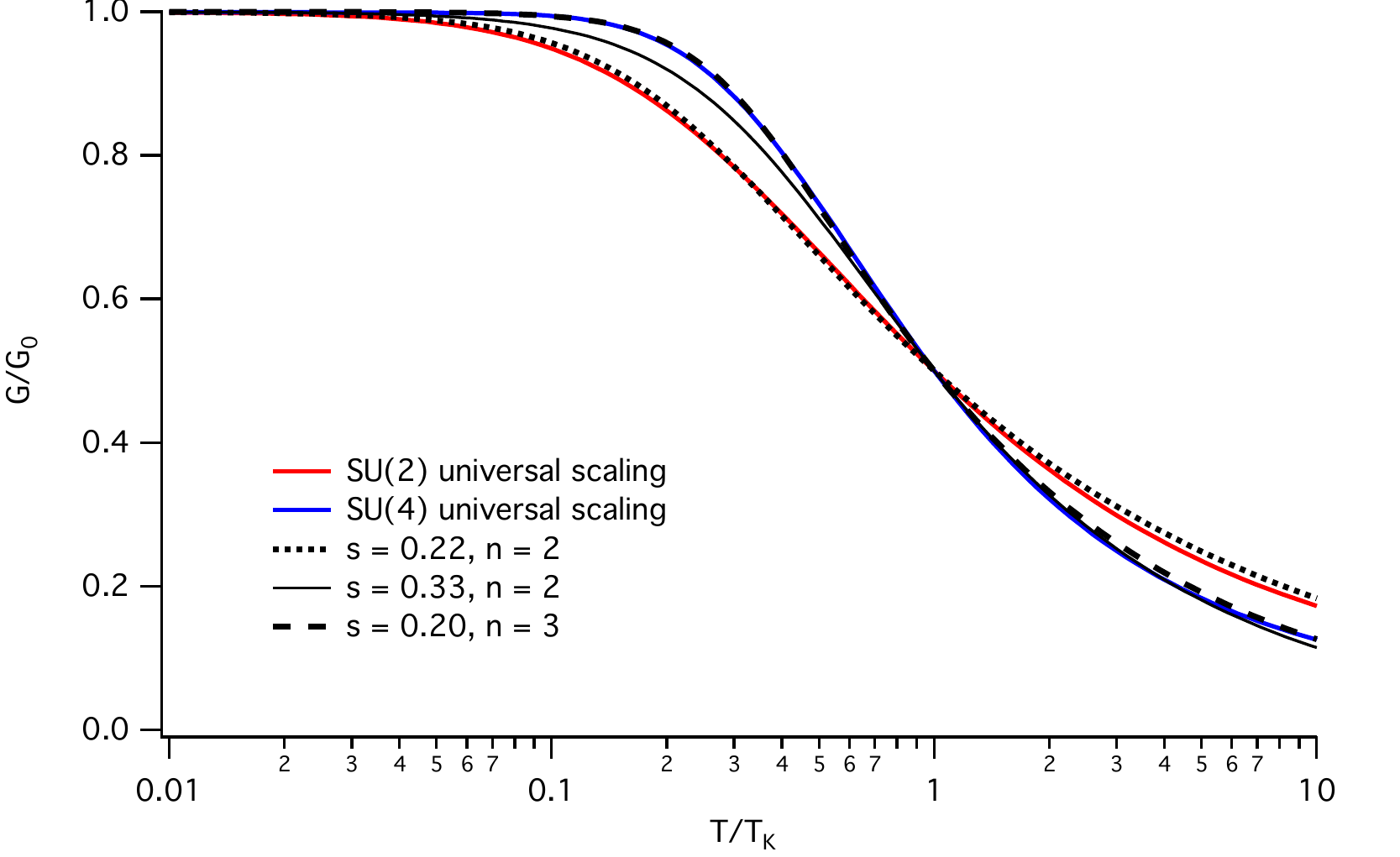}
\caption{Universal SU(2) (red) and 1/4-filling SU(4) (blue) scaling curves for the conductance as a function of temperature. $T_{K SU(2)}$ and $T_{K SU(4)}$ are both defined such that $G/G_0 = 0.5$. Also shown are empirical fits in the form of (\ref{su2form}): $s = 0.22$, $n = 2$ describes SU(2) (black dotted);  $s = 0.33$, $n = 2$ best approximates the SU(4) form without changing $n$ (solid black); $s = 0.20$, $n = 3$ provides a good approximation of the SU(4) form. \label{empirical}}
\end{center}
\end{figure}

In particular, the leading-order temperature dependence of (\ref{su2form}) is quadratic by design at $T \ll T_K$ in order to describe SU(2) Kondo scaling, but conformal field theory predicts the SU(4) Kondo state to have a leading-order cubic temperature dependence at $T \ll T_K$, despite retaining a Fermi liquid character (normally associated with quadratic dependence) \cite{lehur}. Therefore, both parameters $s$ and $n$ must be changed to expect a nice agreement for $T \lesssim T_K$, where the empirical form is designed to apply. Fig. \ref{empirical} shows how $s=0.22$, $n=2$ describes SU(2) universal scaling in the crossover regime. Changing $s$ alone is seen to be insufficient to describe the SU(4) universal scaling especially for temperatures $T < T_K$, where the fitting is most sensitive. However, a good fit to the SU(4) universal scaling may be obtained with $s=0.20$, $n=3$. We must emphasize that although (\ref{su2form}) provides an accurate fitting in the full crossover region, it fails at temperatures $T \gg T_K$, where it does not reproduce the well-known logarithmic behavior characteristic of the Kondo problem.

From our experiences with analyzing the experimental data in this paper, empirical forms must be used with great care and supported by other methods. A blind application to our data would yield spurious conclusions, owing to the saturation at $T = 40$~mK. Also, as can be seen from the NRG results for our device, there are some expected deviations from the universal scaling, particularly at $T > T_K$, where the empirical forms become less accurate.

\begin{figure}
\begin{center}
\includegraphics[width=6in]{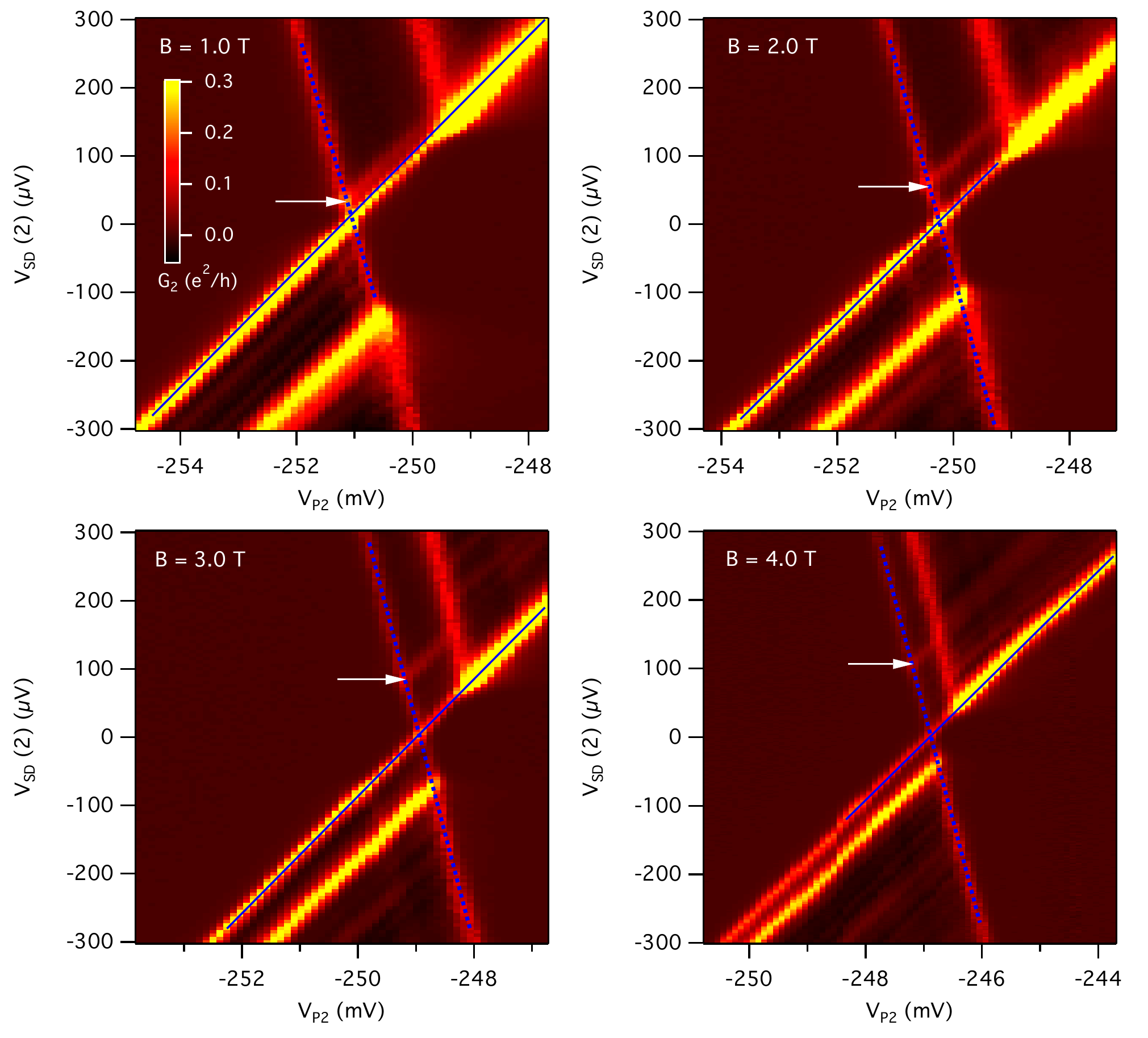}
\caption{Conductance $G_2$ as a function of source-drain bias $V_{SD} (2)$ across dot 2 and gate voltage $V_{P2}$, at in-plane magnetic fields of $B = $ 1.0 T (top-left), $B = $ 2.0 T (top-right), $B = $ 3.0 T (bottom-left), and $B = $ 4.0 T (bottom-right). The color scale is fixed for all four values of magnetic field, which are labeled in the upper-left of each plot. Blue solid lines correspond to the alignment of the source lead Fermi energy with the ground state, and blue dotted lines correspond to alignment of the drain lead Fermi energy with the ground state. White arrows denote where $V_{SD} (2)$ is read off to extract the splitting. \label{gfactor}}
\end{center}
\end{figure}

\section{$g$-factor calibration}

The Zeeman energy $E_Z$ is related to the magnetic field $B$ by $E_Z \equiv |g| \mu_B B$, where $\mu_B$ is the Bohr magneton and g is the g-factor. Among GaAs/AlGaAs heterostructures, the g-factor can vary considerably, and so we calibrate in situ for our device by looking for a Zeeman splitting in the bias spectroscopy as we vary an in-plane magnetic field. Fig. \ref{gfactor} displays conductance through dot 2, demonstrating the Zeeman splitting. A splitting is seen to emerge by $B = 1.0$~T, though the exact splitting is not resolved owing to the width of the level. As the field is increased, we can extract the splitting by reading off the value of $V_{SD (2)}$ above which the source-drain voltage drop is large enough to allow for inelastic spin flip scattering processes. From this value, any offset for true zero bias is then subtracted (usually a few $\mu$V or less). Table \ref{gfacttable} summarizes the extracted splittings and corresponding g-factors. We find $|g|$ consistent with that of bare GaAs, $|g|=0.44$, and take this value in calculating $E_Z$ for given $B$.

\begin{table}
\begin{center}
\begin{tabular}{|c|c|c|}
\hline
Magnetic field (T) & Splitting ($\mu$eV) & $|g|$ \\
\hline
1.0 & --- & --- \\
2.0 & 51 & 0.44\\
3.0 & 80 & 0.46\\
4.0 & 104 & 0.45\\
\hline
\end{tabular}
\caption{ Approximate spin state splittings and corresponding g-factors as a function of magnetic field.  \label{gfacttable}}
\end{center}
\end{table}


\section{Bias spectroscopy at $N_e = 1$ LBTP}

\begin{figure}[p]
\begin{center}
\includegraphics[width=6in]{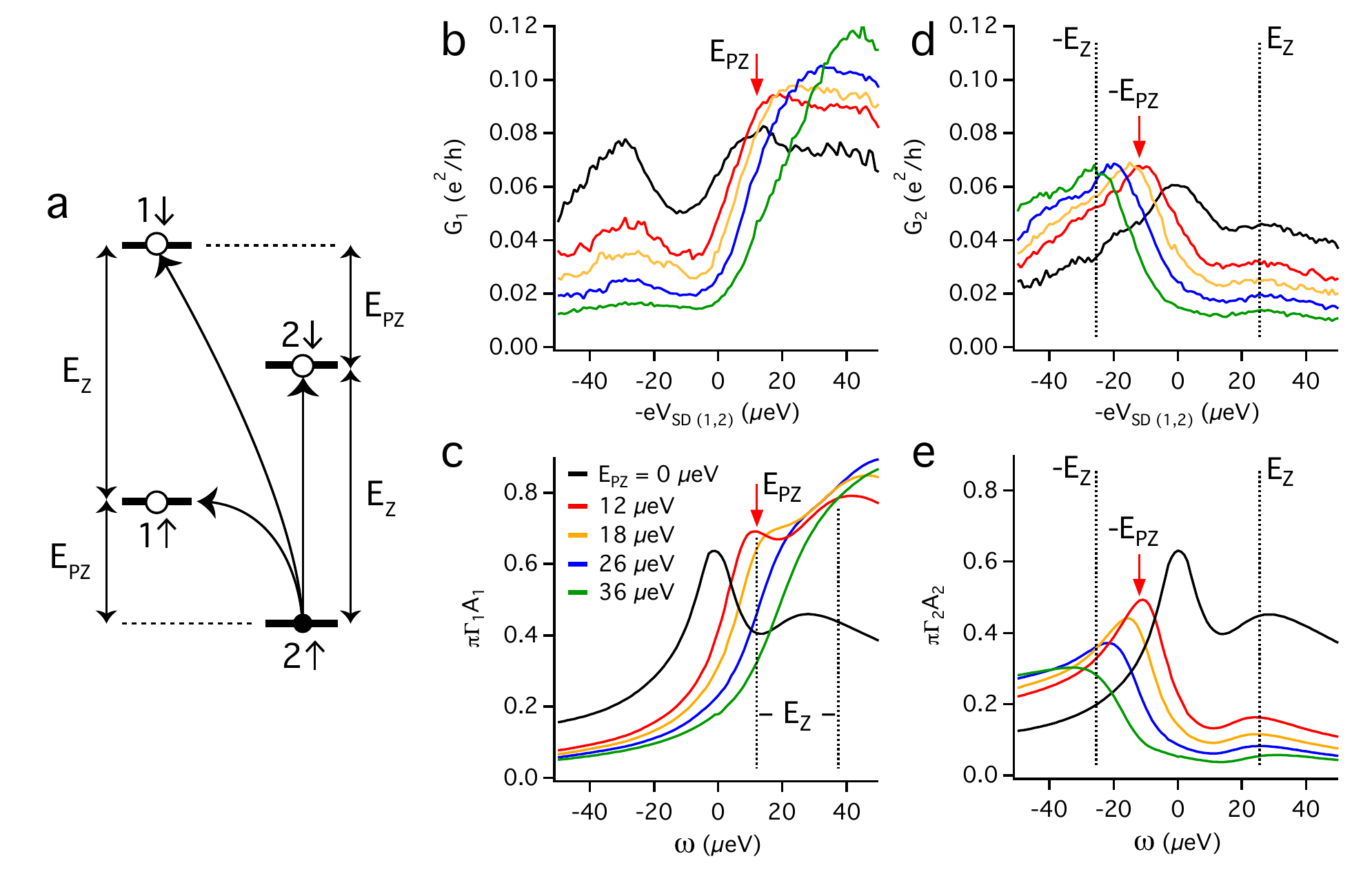}
\caption{
{\bf (a)} Inelastic transitions between Zeeman-split states of dot 1 and dot 2 at an $N_e =1$ LBTP.
{\bf (b)} Experimental conductance $G_1$ for dot 1 in a 1.0 T Zeeman field. The five traces correspond to different values of $E_{PZ}$, with $E_{PZ} > 0$ meaning dot 2 is favored to hold the unpaired electron.
{\bf (c)} Calculated spectral function $A_1$ for dot 1. 
{\bf (d)} Experimental conductance $G_2$ for dot 2.
{\bf (e)} Calculated spectral function $A_2$ for dot 2. 
For all panels, $\Gamma_1, \Gamma_2 \approx 0.04$~meV. $\Gamma_{1S}$ and $\Gamma_{2S}$ were both tuned to be $\sim$ 2--3\% of $\Gamma_{1D}$ and $\Gamma_{2D}$, respectively, such that the biased leads probe the equilibrium local density of states on their respective dot. The bias is applied to both dots simultaneously. The parameters used for the calculations were $T = 40$~mK, $B = 1$~T, $U_1 = 1.2$~meV, $U_2 = 1.5$~meV, $U = 0.1$~meV, $\Delta_1 = 0.017$~meV, $\Delta_2 = 0.0148$~meV. Note that $\alpha_1=\alpha_2=1$ serve only as normalization factors in the calculation. The $\epsilon_1, \epsilon_2$ used are in Table \ref{epsilons}.
\label{ne1LBTP}}
\end{center}
\end{figure}

Fig. \ref{ne1LBTP} shows the orbital state-resolved bias spectroscopy and calculated spectral functions at an $N_e = 1$ LBTP, in a 1.0 T Zeeman field. The spectral functions shown are the same as those shown in Fig. 4, up to the particle-hole transformation that was applied to describe the $N_e = 3$ LBTP. By considering the cartoon of Fig. \ref{ne1LBTP}a, and identifying each electron-like process with a corresponding hole-like process in Fig. 4a, the relationship between the $N_e = 1$ LBTP and $N_e = 3$ LBTP becomes clearer. We again consider $\omega$ and $-eV_{SD}$ as equivalent.

In dot 2, all of the expected features are observed (Fig. \ref{ne1LBTP}d): a weak peak at $\omega = E_Z$, a peak (threshold) that tracks with $E_{PZ}$ for $E_{PZ} < E_Z$, and a purely orbital Kondo peak at $\omega = 0$ for $E_{PZ} = 0$. The overall shapes of the curves are in rough qualitative agreement with the spectral functions in Fig. \ref{ne1LBTP}e, although the relative peak heights may differ.

\begin{table}
\begin{center}
\begin{tabular}{|c|c|c|}
\hline
$E_{PZ}$ (meV) & $\epsilon_1$ (meV) & $\epsilon_2$ (meV) \\
\hline
0 & -0.06333 & -0.06167 \\
0.012 & -0.05667 & -0.06833 \\
0.018 & -0.05387 & -0.07113 \\
0.026 & -0.04966 & -0.07534 \\
0.036 & -0.0445 & -0.0805 \\
\hline
\end{tabular}
\caption{Parameters $\epsilon_1$ and $\epsilon_2$ used for each value of experimental $E_{PZ}$ in Fig. 4 and Fig. \ref{ne1LBTP}. \label{epsilons}}
\end{center}
\end{table}

However, in dot 1 (Fig. \ref{ne1LBTP}b), the purely orbital Kondo peak at $\omega = 0$ for $E_{PZ} = 0$ is obscured by poorly understood background conductance at positive $\omega$. Additionally, an unexpected feature is observed at $\omega = -30$~$\mu$V that does not track with $E_{PZ}$. It is tempting to suggest that the LBTP being measured is actually a (1,1)/(2,0) LBTP. In this interpretation, both dots could hold an unpaired electron, and both dots should exhibit a peak at $\omega = \pm E_{Z}$. In other words, the spectral functions for both dots should look similar to Fig. \ref{ne1LBTP}e, with $\omega \rt -\omega$ for dot 1. However, the increasing conductance at positive $\omega$ in Fig. \ref{ne1LBTP}b is in qualitative agreement with Fig. \ref{ne1LBTP}c, and would not be expected in this alternate explanation. Additionally, our ability to maintain electron occupation number assignments is supported by Fig. \ref{surveyB0}. Therefore, the unexpected feature is instead likely associated with a low-lying excited state.

\section{Technical details}

\subsection{Electronics \label{electronics}}

For the data taken in Fig. 1b and 4 of the paper, custom current amplifiers designed by Y. Chung of Pusan National University (early version of that which is presented in \cite{kretinin}) were used in place of commercial Ithaco / DL Instruments 1211 current amplifiers, which have been previously employed in our measurement setup \cite{potok}. The custom amplifiers are crucial to this experiment in that the input offset voltage of the current amplifiers must remain stable over a period of days to avoid applying an uncontrolled source-drain bias across the dot. 
	Over a continuous interval of 2.8 days, the standard deviation of the input offset voltage was measured to be 1.0 $\mu$V for the amplifier attached to dot 1, and 0.6 $\mu$V for the amplifier attached to dot 2. The amplifiers were characterized in the same locations where they were used for measurement, as no active temperature control of the amplifiers was performed during measurement or characterization.

\subsection{Magnetic field calibration}

\begin{figure}[h!]
\begin{center}
\includegraphics[width=3in]{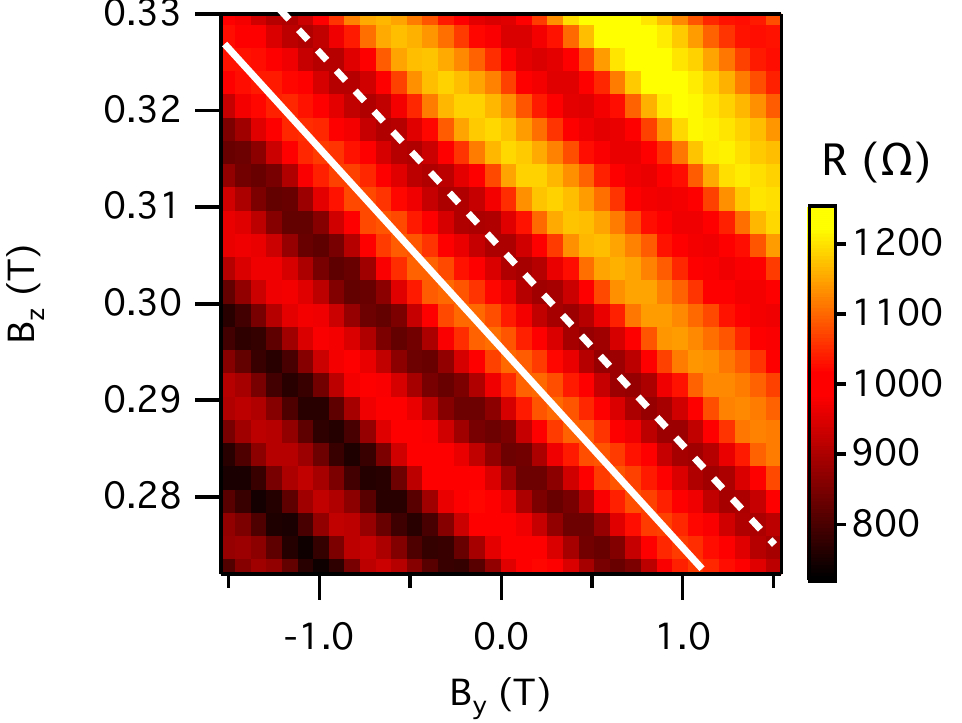}
\caption{Four-wire resistance as a function of the y-axis (in-plane) and z-axis (perpendicular) magnetic fields. The slopes of the solid white and dashed white lines are $m = -0.0206$ and $m = -0.0203$, respectively. This corresponds to a 1.2$^\circ$ misalignment between the y-axis field and the plane of the sample.\label{shubnikov}}
\end{center}
\end{figure}

Because of small but uncontrolled sample tilt with respect to axes defined by the two-axis magnet in our experimental dewar, energizing only the in-plane coil will give rise to a perpendicular component as seen by the sample, and vice versa. To apply a magnetic field precisely in the plane of the sample, as is done in Fig. 4, we calibrate in situ using a four-wire current-biased measurement of Shubnikov-de Haas oscillations in resistance, as a function of both the nominally perpendicular field $B_z$ and nominally in-plane magnetic field $B_y$.

Fig. \ref{shubnikov} shows the Shubnikov-de Haas oscillations observed near a perpendicular magnetic field of 0.3 T, and how they track with an added in-plane field. The geometry of the 2DEG mesa is not well defined, so both even and odd components of magnetoresistance contribute to the measured resistance. The observed stripes correspond to a constant perpendicular field. The slope of the stripes gives a compensation factor such that any perpendicular component introduced by the in-plane magnetic field may be cancelled out by application of an added perpendicular field to within a few percent.

Even an applied field in the plane of the sample will subtly modify orbital states because of the finite extent of the electronic wavefunctions normal to the plane, an effect we neglect in our analysis.

\subsection{Bias spectroscopy}

To apply and maintain a particular $E_{PZ}$ while changing the applied source-drain biases $V_{SD1(2)}$ across dot 1 (2) requires some care. Gates P1 and P2 as well as leads S1 and S2 all have capacitances to both dot 1 and dot 2. These capacitances must all be characterized every time the W gates or magnetic field are changed. Once the capacitances are known, electrostatic gating of the dots by the biased source leads may be compensated by changes in $V_{P1}$ and $V_{P2}$. Further details have been published previously \cite{Amasha2013}.